\documentclass[manuscript]{aastex61}

\def\VEC#1{\mbox{\boldmath $#1$}}

\received{2019 June 4}
\revised{2019 June 12}
\accepted{2019 June 27}

\shorttitle{Method and test calculations of 1D Force-free magnetodynamics}
\shortauthors{Koide and Imamura}

\begin{document}

\title{Analytical and Numerical Methods and Test Calculations of One-Dimensional 
Force--Free Magnetodynamics on Arbitrary Magnetic Surfaces
across Horizons of Spinning Black Holes}

\correspondingauthor{Shinji Koide}
\email{koidesin@kumamoto-u.ac.jp}

\author{Shinji Koide}
\affiliation{Department of Physics, Faculty of Science, Kumamoto University,
2-39-1, Kurokami, Kumamoto, 860-8555, JAPAN}

\author{Tomoki Imamura}
\affiliation{Algorithm Laboratory,
2-6-14, Ebisu-minami, Shibuya-ku, Tokyo, 150-0022, JAPAN}



\begin{abstract}
Numerical simulations of the force-free magnetodynamics (FFMD) of 
the electromagnetic field
around a spinning black hole are useful to investigate
the dynamic electromagnetic processes around a spinning black hole, such as
the emergence of the Blandford--Znajek mechanism.
To reveal the basic physics of magnetic fields around a black hole
through the dynamic process,
we use one-dimensional (1D) FFMD along the axisymmetric magnetic surface, 
which provides 
a relatively simple, sufficiently precise, and powerful tool to analyze
the dynamic process around a spinning black hole.
We review the analytic and numerical aspects of 1D FFMD
for an arbitrary magnetic surface around a black hole.
In addition, 
we also show some numerical simulation test results for three types of magnetic surfaces
at the equatorial plane of the black hole.
\end{abstract}

\keywords{ black hole physics --- Galaxy: nucleus --- magnetic fields 
 --- methods: analytical --- methods: numerical --- plasmas }



\section{Introduction} \label{sec2}
Recently, images of the supermassive black hole shadow in the center
of the giant elliptical galaxy M87, whose central core emits a powerful 
relativistic jet almost toward us \citep{biretta99}, 
have been reconstructed by the team
of the Event Horizon Telescope (EHT) collaboration \citep{eht19a}.
To extract information on the dynamics of the plasma, 
the black hole's gravitational field, and the black hole itself,
it is necessary to develop fully general relativistic models of the accretion flow, 
associated winds and relativistic jets, and the emission properties 
of the plasmas.
The general relativistic models are also required to understand the formation of
relativistic jets from stellar--mass black holes, such as 
microquasars (black hole binaries; \citealt{mirabel94,tingay95}) and
gamma-ray bursts (GRBs; \citealt{kulkarni99,ligo17})
and gravitational waves emitted from merging
stellar--mass black holes \citep{abbott16}.
The most common approach to dynamical relativistic source modeling uses
the ideal general relativistic magnetohydrodynamic (GRMHD) approximation.
Over the last decades, a number of GRMHD codes have been developed
and applied to a large variety of astrophysical scenarios
\citep{koide98,koide99,koide00,koide02,koide06,gammie03,koide03,
mckinney06,delzanna07,mckinney09,mckinney13,radice13}.
The EHT team also found that images produced from GRMHD simulations 
with general relativistic ray-tracing calculations \citep{eht19b,porth19}
are consistent with images of the asymmetric ring seen in the EHT data.
From a comparison between GRMHD simulations and
EHT images, the EHT collaboration team concluded 
that the asymmetry in brightness in the ring can be explained 
in terms of relativistic beaming of the emission from a plasma rotating
close to the speed of light around a black hole spinning clockwise.
At the same time, in those models that produce the relativistic jet of M87,
the jet is powered by extraction of black hole spin energy through
the Blandford-Znajek mechanism \citep{blandford77,mckinney06}.
The relativistic jet is anchored to the ``funnel" region near the polar
axis where low angular momentum material will be swallowed up by
the black hole. The strong magnetic fields that permeate the black hole
make the Blandford-Znajek mechanism work.
The ensuing near-vacuum and magnetic dominance are difficult for
GRMHD simulations to handle. In such a region near the polar axis, 
the force-free condition becomes a good approximation.

To investigate dynamic processes of extremely strong magnetic fields
around a spinning black hole, two-dimensional calculations of 
the force-free magnetodynamics (FFMD)
have been performed
\citep{komissarov01,komissarov02,komissarov04}.
Such simulations showed the emergence
of energy radiation from a spinning black hole via
an axisymmetric magnetic field.
However, sufficient analysis of the emergence of energy radiation
has not yet been shown because we have no analytic solution for a two-dimensional
electromagnetic field around a spinning black hole to compare
with the numerical results.
When we assume that an axisymmetric magnetic surface is fixed around a black hole,
we can reduce FFMD to one--dimensional FFMD (1D FFMD), where
we can obtain the analytic solutions of the steady-state force-free
fields to compare the numerical results.
To investigate the fundamental physics of the magnetic fields dynamically,
we can use numerical simulations of 1D FFMD for 
the axisymmetric magnetic surfaces across the black hole effectively.
In this paper, we review a method of 1D FFMD for an arbitrary fixed 
axisymmetric
magnetic surface around a black hole. The 1D FFMD has the following advantages,
\begin{itemize}
\item Developing the numerical calculation code is easy and 
small computer resources are sufficient to run the code.
\item Analyses of the numerical calculations are relatively easy.
\item An analytic solution of the steady state can be given for an arbitrary
magnetic surface around a rapidly rotating black hole
(Section \ref{anasol4stestafff}).
\end{itemize}
However, 1D FFMD has some disadvantages.
\begin{itemize}
\item An axisymmetric stationary magnetic surface should be assumed.
The results strongly depend on the configuration of the magnetic surfaces.

\item We cannot treat the interaction between the magnetic surfaces and
should neglect energy transport and wave propagation across the magnetic surfaces.
\end{itemize}
It is noted that the energy transport and wave propagation
across the magnetic surface are negligible near the horizon. 
Therefore, we expect that the essential physics
of force-free fields around a black hole
could be seized via 1D FFMD.


We have performed 1D FFMD numerical simulations of a magnetic field
around a spinning black hole at its equatorial plane to investigate
the dynamic process of the energy extraction from the black hole
via the magnetic field \citep{koide18,imamura19}. 
In the case of the radial magnetic surface around an equatorial plane,
the Poynting flux emerges from the ergosphere, and the region of the
finite Poynting flux expands toward infinity like a tsunami
\citep{koide18}. On the other hand, in cases other than the radial
magnetic surface, such as incurvature- or excurvature-flared magnetic surfaces,
the Poynting flux emerges around the ergosphere transiently at a very early stage
of the simulation: however,
the finite Poynting flux region spreads outward very slowly or 
is reduced and eventually vanishes 
\citep{imamura19}. We explained the drastic difference between
these cases based on the analytic solution
of the steady-state force-free fields.

In this paper, we generalize the 1D FFMD equations to perform numerical simulations
of the force-free field along an arbitrary magnetic surface.
In addition, we show the analytic solutions of the steady-state field along
an arbitrary magnetic surface. Test calculations are shown for cases of magnetic surfaces
at the equatorial plane with three types of shapes.

We review generalized 1D FFMD equations for arbitrary
magnetic surfaces in Section \ref{sec2}.
In Section \ref{anasol4stestafff}, 
we provide analytical solutions
of the 1D steady-state force-free fields along an arbitrary
axisymmetric magnetic surface around a spinning black hole.
In Section \ref{sec4}, we present the numerical test results of 1D FFMD
for cases of three types of magnetic surfaces along 
the equatorial plane.
A summary and discussions are shown in Section \ref{secdiscussion}.


\section{Review of FFMD \label{sec2}}

\subsection{Covariant form of FFMD equations}

The FFMD equations are based on the Maxwell equations with
force-free conditions \citep{komissarov01,komissarov04}.
Here we assume the magnetic field is so strong that we can
neglect the plasma inertia. We also assume that the electric resistivity
and self gravity of the electromagnetic field vanish.
The line element in the spacetime $x^\mu = (t, x^1, x^2, x^3)$
is written by $ds^2 = g_{\mu \nu} dx^\mu dx^\nu$.
Here Greek subscripts like $\mu$ and $\nu$ run from 0 to 3,
while Roman subscripts like $i$ and $j$ run from 1 to 3.
We employ the natural unit system, where the speed of light $c$,
gravitational constant $G$, and magnetic permeability and electric
permittivity in the vacuum $\mu_0$, $\epsilon_0$ are unity. 
We also often set the black hole mass unity, $M=1$.
The covariant forms of the Maxwell equations are
\begin{eqnarray}
& \displaystyle \nabla _\mu F^{\mu \nu} = \frac{1}{\sqrt{-g}} \frac{\partial}{\partial x^\mu} 
\left ( \sqrt{-g} F^{\mu \nu} \right )= - J^\nu ,
\label{eqam} \\
& \displaystyle  {\nabla_\mu} {^\ast}F^{\mu \nu} = \frac{1}{\sqrt{-g}} 
\partial _\mu ( \sqrt{-g}  {^\ast}F^{\mu \nu} ) = 0 
\label{eqfa_dual}
\end{eqnarray}
where $\nabla_\mu$ is the covariant derivative,
$F_{\mu \nu}$ is the electromagnetic field tensor,
${^\ast}F_{\mu \nu} = \epsilon^{\mu \nu \rho \sigma} F_{\rho \sigma}/2$ is the 
dual tensor of $F_{\mu \nu}$, and $J^\mu =(\rho_{\rm e}, J^1, J^2, J^3)$
is the four-current density ($\rho_{\rm e}$ is the electric charge density).
Here $\epsilon^{\mu \nu \rho \sigma} = \eta^{\mu \nu \rho \sigma}/\sqrt{-g}$
is the Levi--Civita tensor, $g = {\rm det} (g_{\mu \nu})$ 
is the determinant of $(g_{\mu \nu})$, and $\eta^{\mu \nu \rho \sigma}$
is the totally asymmetric symbol
\footnote{$\eta^{\mu \nu \rho \sigma} = 1$
if the order $[\mu \nu \rho \sigma]$ is an even permutation of $[0123]$,
$\eta^{\mu \nu \rho \sigma} = -1$
if the order $[\mu \nu \rho \sigma]$ is an odd permutation of $[0123]$,
and $\eta^{\mu \nu \rho \sigma} = 0$
if $\mu, \nu, \rho, \sigma$ are not all different.}.
We use the electric field $E_\mu$ and magnetic field $B^\mu$ given by
\begin{eqnarray}
E_\mu = F_{\mu 0} , \verb!   !
B^\mu = {^*}F^{0 \mu} = \frac{1}{2} \epsilon^{0 \mu \rho \sigma} F_{\rho \sigma}.
\end{eqnarray}


The Maxwell equations read the conservation
law of the electromagnetic energy and momentum,
\begin{equation}
\nabla_\mu T^{\mu \nu} = - f_\nu^{\rm L},
\label{eqenmo}
\end{equation}
where $T^{\mu \nu} = {F^\mu}_\sigma F^{\nu \sigma} 
- \frac{1}{4}g^{\mu \nu} F^{\lambda \kappa} F_{\lambda \kappa}$
is the four-electromagnetic energy-momentum tensor and
$f^\nu_{\rm L} = J_\mu F^{\mu \nu}$ is the four-Lorentz force density.
In the case of the force-free condition $f_\nu^{\rm L} = 0$, 
Eq. (\ref{eqenmo}) yields
\begin{equation}
\nabla_\mu T^{\mu \nu} = 0.
\label{eqenm0}
\end{equation}
To perform FFMD numerical simulations, we use 
Eq. (\ref{eqenmo}) instead of Eq. (\ref{eqam})

\subsection{Degeneracy of force-free electromagnetic field}

The degeneracy of the force-free electromagnetic field 
is imposed on the FFMD supplementarily. This degeneracy is due to the condition
of the magnetospheric plasma, where the charged particles are plentiful
enough to support a strong electric current and screen the electric field
\citep{goldreich69}. To derive the degeneracy, we consider the low-inertia
limit of relativistic magnetohydrodynamics (RMHD). 
The relativistic Ohm's law is
\begin{equation}
u^\nu F_{\mu\nu} = \eta (J_\mu - \rho_{\rm e}^\prime u_\mu)
\label{ohmlaw4fm}
\end{equation}
where $u^\mu$ is the four-velocity of the plasma, $\eta$ is the resistivity,
$\rho_{\rm e}^\prime$ is the proper electric density.
Eq. (\ref{ohmlaw4fm}) yields
\begin{equation}
E_{\tilde{i}} + \epsilon_{ijk} v^{\tilde{j}} B^{\tilde{k}} 
= \frac{\eta}{\gamma}_{\rm L} (J_{\tilde{i}} - \rho_{\rm e}^\prime u_{\tilde{i}})
= \eta \acute{J}_{\tilde{i}}
\end{equation}
where $\gamma_{\rm L} = u^{\tilde{0}}$,
$v^{\tilde{i}} = u^{\tilde{i}} /\gamma_{\rm L}$, and
$\acute{J}_{\tilde{i}} \equiv \frac{1}{\gamma_{\rm L}} (J_{\tilde{i}}
- \rho^\prime_{\rm e} u_{\tilde{i}})$ is similar to the current density
observed by the plasma rest frame but not exactly.
We have
\begin{eqnarray}
& \displaystyle  ^\ast F^{\rho \sigma} F_{\rho \sigma} = 
^\ast F^{\tilde{\rho} \tilde{\sigma}} F_{\tilde{\rho} \tilde{\sigma}} = 
-4 E_{\tilde{i}} B^{\tilde{i}} = -4 \frac{\eta}{\gamma_{\rm L}} 
(J_{\tilde{i}} - \rho_{\rm e}^\prime u_{\tilde{i}}) B^{\tilde{i}}, 
\label{sfrsfrs} \\
& \displaystyle  F^{\rho \sigma} F_{\rho \sigma} = 
F^{\tilde{\rho} \tilde{\sigma}} F_{\tilde{\rho} \tilde{\sigma}} = 
2 ( B^{\tilde{i}} B_{\tilde{i}} - E^{\tilde{i}} E_{\tilde{i}} )
= 2 \left [
\frac{1}{\gamma^2_{\rm L}} B_{\tilde{i}} B^{\tilde{i}} +
(v_{\tilde{i}} B^{\tilde{i}})^2 - \eta \left \{
\eta J_{\tilde{i}} J^{\tilde{i}} + \frac{2}{\gamma_{\rm L}}
\epsilon_{ijk}  v^{\tilde{i}} J^{\tilde{j}} B^{\tilde{k}}
\right \} \right ]
\label{frsfrs}
\end{eqnarray}
When $\eta$ vanishes, Eqs. (\ref{sfrsfrs}) and (\ref{frsfrs}) yield the degeneracy,
\begin{eqnarray}
^\ast F^{\rho \sigma} F_{\rho \sigma} &= 0 ,
\label{eq2degeneracy} \\
F^{\rho \sigma} F_{\rho \sigma} & > 0 .
\label{eq2magflddom}
\end{eqnarray}
In the case of the resistive plasma, the degeneracy is not always guaranteed.
Because, for example, when $\eta$ and $J_{\tilde{i}} J^{\tilde{i}}$ are finite and 
$u_{\tilde{i}}$ vanishes, $^\ast \, F^{\rho \sigma} F_{\rho \sigma} \ne 0$
from Eq. (\ref{sfrsfrs}). Furthermore, when $B^{\tilde{i}}$ vanishes, Eq. (\ref{frsfrs})
yields $F^{\rho \sigma} F_{\rho \sigma} = - 2 \eta^2 \acute{J}^2 < 0$
where the degeneracy are broken.

\subsection{3+1 formalism of FFMD}

We review the 3+1 formalism of the FFMD equations
derived from the covariant Eqs. (\ref{eqam}), (\ref{eqfa_dual}), 
and (\ref{eqenmo}).
In order to introduce the 3+1 formalism, we use
the local coordinate frame called ``normal observer frame" $x^{\tilde{\mu}}$, 
in which the line element is written by,
\begin{equation}
ds^2 = - d \tilde{t}^2 + \gamma_{ij} dx^{\tilde{i}} dx^{\tilde{j}}.
\label{metricoblminkov}
\end{equation}
Here we treat $\gamma_{ij}=g_{ij}$ as the elements
of a $3 \times 3$ matrix $(\gamma_{ij})$ and $\gamma = {\rm det}(g_{ij})$.
Then $\gamma^{ij}$ indicates the elements of the inverse of the matrix $(g_{ij})$,
that is $\gamma^{ik} \gamma_{kj} = \delta^i_j$.
When we define the lapse function $\alpha$ and shift vector $\beta^i$ as
$g_{i0} = g_{0i} = g_{ij} \beta^j$ (or $\beta_i = g_{i0}$),
$\alpha^2 = - g_{00} + g_{ij} \beta^i \beta^j$, we write the line element as
\begin{equation}
ds^2 = - \alpha^2 + g_{ij} (dx^i + \beta^i dt)(dx^j + \beta^j dt).
\label{metric431}
\end{equation}
Comparing Eqs. (\ref{metricoblminkov}) and (\ref{metric431}), we obtain 
$d \tilde{t} = \alpha dt$,
$dx^{\tilde{i}} = dx^i + \beta^i dt$
and we have $\partial_{\tilde{t}} = \alpha^{-1} ( \partial_t - \beta^i \partial_i)$,
$\partial_{\tilde{i}} = \partial_i$.

This local coordinate system is not always orthonormal in the space.
Then the components of the vectors and tensors in the local reference frame
is not intuitive in general.
In the Boyer--Lindquist (BL) coordinates, the local frame of space is already orthogonal,
$g_{ij} = h_i^2 \delta_{ij}$,
and then we can orthonormalized easily as
\begin{eqnarray}
ds^2 = \eta_{\mu \nu} d {x}^{\hat{\mu}} d {x}^{\hat{\nu}} 
= - d \hat{t}^2 + \sum_{i=1}^3 (d {x}^{\hat{i}})^2
= - d \tilde{t}^2 + \sum_{i=1}^3 h_i h_j \delta_{ij} d {x}^{\tilde{i}} d {x}^{\tilde{j}},
\end{eqnarray}
where we set $d \hat{t} = d \tilde{t}$, 
$d x^{\hat{i}} = h_i dx^{\tilde{i}}$. 
We have $\partial_{\hat{t}}=\partial_{\tilde{t}}$, 
$\partial_{\hat{i}} = h_i^{-1}\partial_{\tilde{i}}$.
Here $\eta_{\mu \nu}$ is the metric of Minkowski space-time.
This orthonormal local reference frame $x^{\hat{\mu}}$ 
is called the ``zero-angular-momentum observer" (ZAMO) frame.
%


The 3+1 formalism of the Maxwell equations are
\begin{eqnarray}
& \displaystyle  \frac{1}{\sqrt{\gamma}} \frac{\partial}{\partial x^i} (\sqrt{\gamma} B^{\tilde{i}}) = 0
\label{gauss2mag} ,\\
& \displaystyle  \frac{\partial}{\partial t} B^{\tilde{i}}
= - \epsilon^{{i}{j}{k}} 
\frac{\partial}{\partial x^j} \left [
\alpha ( E_{\tilde{k}} - {\epsilon}_{kpq} N^p B^{\tilde{q}})
\right ]
\label{faraday} , \\
& \displaystyle  \frac{1}{\sqrt{\gamma}} \frac{\partial}{\partial x^i} (\sqrt{\gamma} E^{\tilde{i}}) 
= \tilde{\rho}_{\rm e}
\label{gauss2ele} , \\
& \displaystyle  \frac{\partial}{\partial t} E^{\tilde{i}}
+ \alpha (J^{\tilde{i}} + \tilde{\rho}_{\rm e} N^i)
= \epsilon^{{i}{j}{k}} \frac{\partial}{\partial x^j}
[\alpha (B_{\tilde{k}} + {\epsilon}_{kmn} N^m E^{\tilde{n}})]
\label{ampere}
\end{eqnarray}
where $\epsilon^{ijk} = \alpha \epsilon^{0ijk}= \gamma^{-1/2} \eta^{0ijk}$
is the three-dimensional Levi--Civita tensor,
$N^i = - \beta^i/\alpha$. Note that $N^\mu = (1/\alpha, N^i)$ is the
four-velocity of the normal observer.

The 3+1 formalism of the force-free condition is
\begin{eqnarray}
\tilde{J}^i \tilde{E}_i = 0, \hspace{0.7cm}
\tilde{\rho}_{\rm e} \tilde{E}_i + {\epsilon}_{ijk} \tilde{J}^j = 0.
\label{forcefreecon}
\end{eqnarray}

The conservation laws of energy and momentum of the electromagnetic field are
given by
\begin{eqnarray}
& \displaystyle   \frac{\partial}{\partial t} \tilde{u}
+ \frac{1}{\sqrt{\gamma}} \frac{\partial}{\partial x^i}
[\alpha \sqrt{\gamma} (\tilde{S}^i + {N}^i \tilde{u})]
+ \frac{\partial \alpha}{\partial x^i} \tilde{S}^i
\nonumber \\
& \displaystyle  + \left [ \gamma_{jk} \frac{\partial}{\partial x^i} (\alpha N^k) 
+ \frac{1}{2} \alpha N^k \frac{\partial}{\partial x^k} \gamma_{ij}
\right ] T^{\tilde{i}\tilde{j}}
= - \alpha \tilde{J}^i \tilde{E}_i
\label{3+1enem} , \\
& \displaystyle   \frac{\partial}{\partial t} \tilde{S}_i
+ \frac{1}{\sqrt{\gamma}} \frac{\partial}{\partial x^k}
[ \alpha \sqrt{\gamma} (T^{\tilde{k}}_{\tilde{i}} + {N}^k \tilde{S}_i) ] 
+ \frac{\partial \alpha}{\partial x^i} \tilde{u}
\nonumber \\
& \displaystyle  + \frac{\partial}{\partial x^i} (\alpha N^k) \tilde{S}_k
- \frac{1}{2} \frac{\partial}{\partial x^i} \gamma_{jk} T^{\tilde{j} \tilde{k}}
= - \alpha (\tilde{\rho}_{\rm e} \tilde{E}_i + {\epsilon}_{ijk} \tilde{J}^j
\tilde{B}^k)
\label{3+1moem}
\end{eqnarray}
where 
$S_{\tilde{i}} = \epsilon_{ijk} E^{\tilde{j}} B^{\tilde{k}}$,
$\tilde{u} = (E^{\tilde{i}} E_{\tilde{i}} + B^{\tilde{i}} B_{\tilde{i}})/2$,
$B_{\tilde{i}} = \gamma_{ij} B^{\tilde{j}}$, and
$E^{\tilde{i}} = \gamma^{ij} E_{\tilde{j}}$
(Appendix \ref{appenda}).

\subsection{Equation of energy conservation \label{appendb}}

Here we derive the equation of energy conservation from Eq. (\ref{eqenmo})
with the Killing vector of temporal boost transformation
$\xi^\mu = (1,0,0,0)$. 
Using the Killing equation $\nabla_\mu \xi_\nu + \nabla_\nu \xi_\mu =0$, we obtain
\begin{equation}
\nabla_\mu (T^{\mu\nu} \xi_\nu) = 
\frac{1}{\sqrt{-g}} \partial_\mu (\sqrt{-g} T^{\mu\nu} \xi_\nu) = 0.
\end{equation}
When we use the energy flux density $S^\mu = T^{\mu\nu} \xi_\nu$ and 
the energy-at-infinity density $e^\infty \equiv S^{\tilde{0}}$, we have the
equation of the energy conservation
\begin{equation}
\frac{\partial}{\partial t} e^\infty + \frac{1}{\sqrt{\gamma}}
\frac{\partial}{\partial x^i} (\alpha \sqrt{\gamma} S^i) =0.
\label{eq2energycons}
\end{equation}

\subsection{Conditions from axisymmetry on flux coordinates}

\begin{figure} 
\begin{center}
\includegraphics[width=8cm]{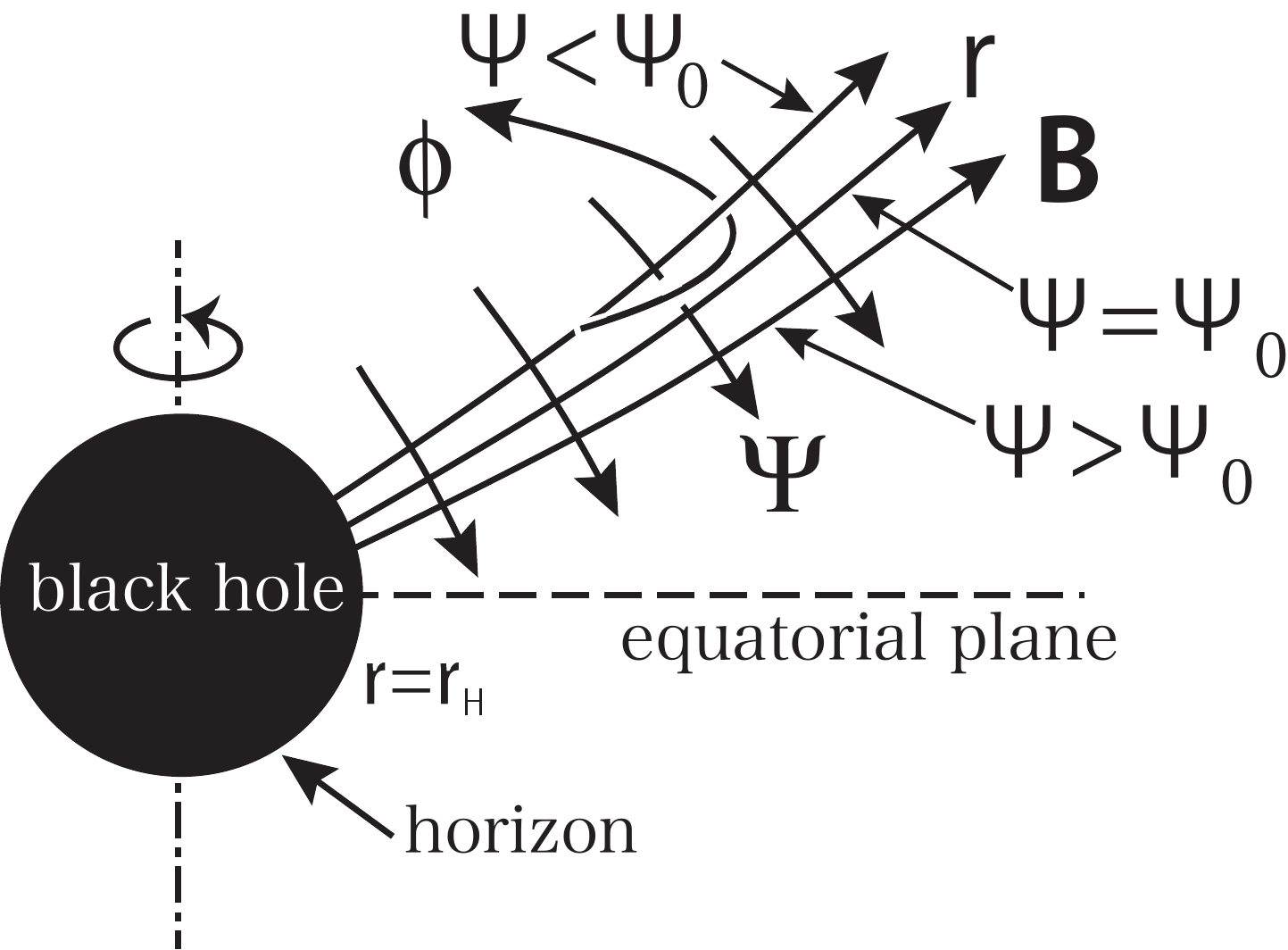}
\caption{Flux coordinates for 1D FFMD simulations
with arbitrary axisymmetric magnetic surface.
We assume that
the electromagnetic field is force-free and axisymmetric. 
\label{naturalcoordin}}
\end{center}
\end{figure}

\begin{figure} 
\begin{center}
\includegraphics[width=10cm]{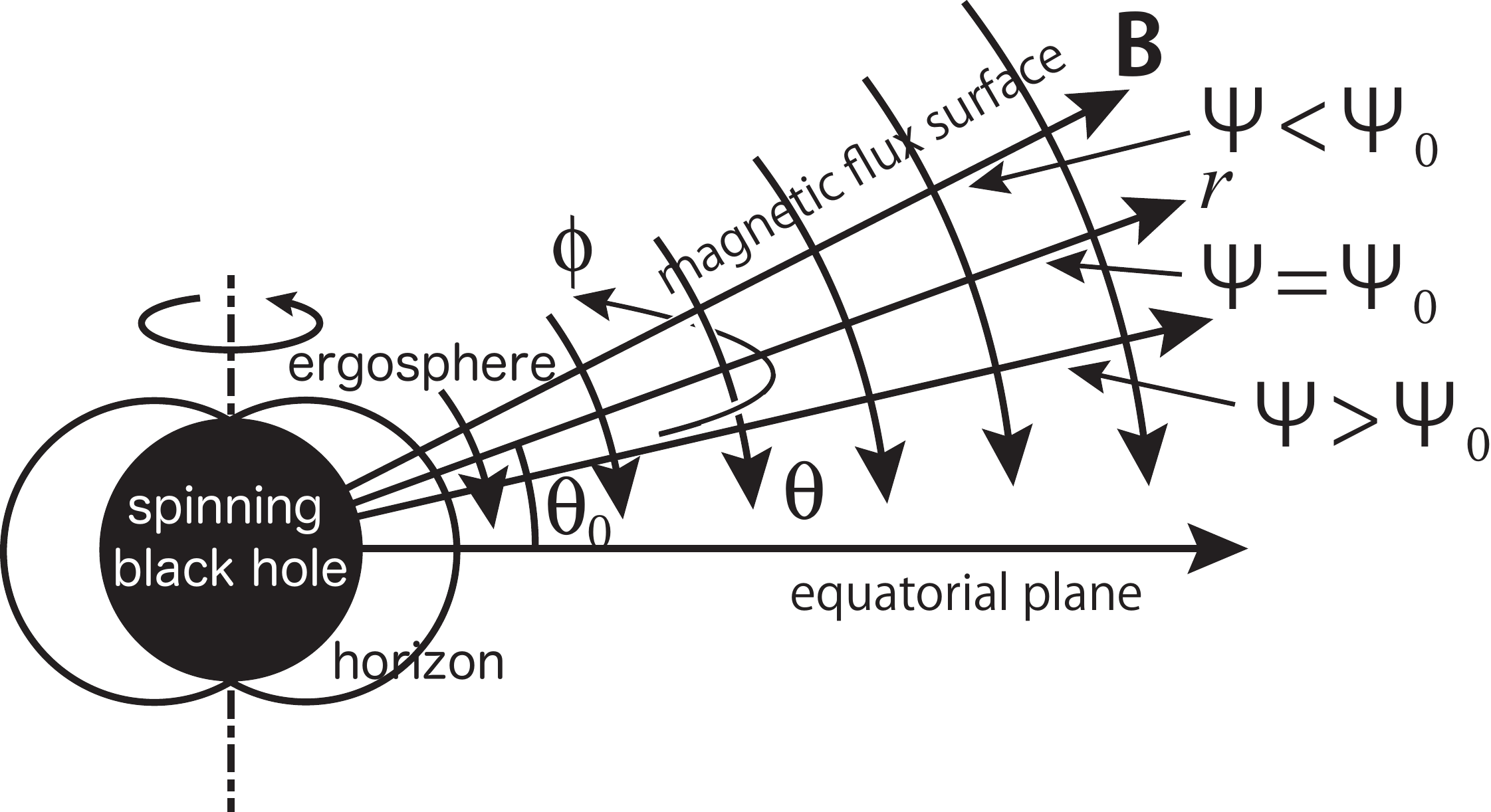}
\caption{Flux coordinates for 1D FFMD simulations
with axisymmetric radial magnetic surface $\Psi = \Psi_0$.
We assume that the electromagnetic field is force-free and axisymmetric. 
\label{ffmd1d}}
\end{center}
\end{figure}

Here we introduce flux coordinates for a given axisymmetric magnetic surface
along an arbitrary axisymmetric surface imaged by Fig. \ref{naturalcoordin}.
Using the vector potential $A^\mu$ for the given axisymmetric 
electromagnetic field, 
the magnetic surface is described by $ \Psi = A_\phi =$ constant.
The flux coordinates are given by $x^\mu = (t, r, \Psi, \phi)$.
Here, $t$, $r$, and $\phi$ are the time, radial, and azimuthal coordinates, respectively.
Using the flux coordinates, we have $B^\Psi = 0$.
For a radial magnetic surface as shown in Fig. \ref{ffmd1d}, 
we use the colatitude coordinate $\theta$ as the coordinate $\Psi - \Psi_0 + \theta_0$,
where the radial magnetic surface is given by $\theta=\theta_0$.
Note that, in this case, we use the unit system so that $\displaystyle 
\frac{\partial \Psi}{\partial \theta}=1$.
In general, the flux coordinates are not orthogonal, while
the coordinates are orthogonal at the equatorial plane.

When we use the flux coordinates $(r, \Psi, \phi)$ 
($B^{\tilde{\Psi}}=0$),
we have the following
three conditions with respect to $E_{\tilde{\phi}}$, $E_{\tilde{r}}$, and 
$J^{\tilde{\Psi}}$ in general.
It is noted that the BL and Kerr--Schild (KS) coordinates
are used for the flux coordinates of radial magnetic surfaces around
the equatorial plane.
In the case of a radial magnetic surfaces, $\Psi$ is directly given by $\theta$
without the addition of $\Psi_0 - \theta_0$.

The $\Psi$-component of Eq. (\ref{faraday}) yields
\[
\frac{\partial B^{\tilde{\Psi}}}{\partial t} = - \frac{1}{\sqrt{\gamma}}
\frac{\partial}{\partial r} 
[\alpha (E_{\tilde{\phi}} + \sqrt{\gamma} N^\Psi B^{\tilde{r}})] = 0 ,
\]
and we find that 
$\alpha (E_{\tilde{\phi}} + \sqrt{\gamma} N^\Psi B^{\tilde{r}})$ is uniform. 
Then we have 
\begin{equation}
E_{\tilde{\phi}} = - \sqrt{\gamma} N^\Psi B^{\tilde{r}} 
\label{cond6magsurcoo}
\end{equation}
because $\alpha E_{\tilde{\phi}}+ \sqrt{\gamma} N^\Psi B^{\tilde{r}}$ 
should vanish at infinity.
The azimuthal component of the force-free condition (Eq. (\ref{forcefreecon})) reads
\[
\tilde{\rho}_{\rm e} E_{\tilde{\phi}} + \epsilon_{\phi\Psi r}
J^{\tilde{\Psi}} B^{\tilde{r}} 
= - \sqrt{\gamma} (J^{\tilde{\Psi}} + \tilde{\rho}_{\rm e} N^\Psi ) B^{\tilde{r}} = 0 .
\]
Then we have 
\begin{equation}
J^{\tilde{\Psi}} = - \tilde{\rho}_{\rm e} N^\Psi
\label{cond6ff}
\end{equation}
because $B^{\tilde{r}}$ is finite.

The degeneracy of the field in FFMD (Eq. (\ref{eq2degeneracy}))
and Eq. (\ref{cond6magsurcoo}) yield
$E_{\tilde{i}} B^{\tilde{i}} = E_{\tilde{r}} B^{\tilde{r}} 
+ E_{\tilde{\phi}} B^{\tilde{\phi}} = B^{\tilde{r}} (E_{\tilde{r}} - \sqrt{\gamma}
N^\Psi B^{\tilde{\phi}}) =0$. Then we have
\begin{equation}
E_{\tilde{r}} = \sqrt{\gamma} N^\Psi B^{\tilde{\phi}}.
\label{cond6degeneracy}
\end{equation}
The relationship given by Eqs. (\ref{cond6magsurcoo}), (\ref{cond6ff}), 
and (\ref{cond6degeneracy})
is important for the following calculations.

\subsection{1-D FFMD equations}

We derive a general form of 1D FFMD equations for an arbitrary 
axisymmetric magnetic surface of the force-free electromagnetic field.
Eqs. (\ref{gauss2mag})-(\ref{ampere}), and (\ref{3+1moem}), 
and (\ref{eq2energycons}) yield the following ``1D" equations:
\begin{eqnarray}
& \displaystyle \frac{1}{\sqrt{\gamma}} \frac{\partial}{\partial r} (\sqrt{\gamma} B^{\tilde{r}}) = 0, 
\hspace{1cm} 
 \frac{\partial}{\partial t} B^{\tilde{r}} = 0 , \label{gauss4b1d}\\
& \displaystyle \frac{\partial}{\partial t} E^{\tilde{\Psi}}
 = - \frac{1}{\sqrt{\gamma}} \frac{\partial}{\partial r}
 [\alpha (B_{\tilde{\phi}} + \sqrt{\gamma} N^r E^{\tilde{\Psi}}
- \sqrt{\gamma} N^\Psi E^{\tilde{r}})] , 
\label{ampere31} \\
& \displaystyle \frac{\partial}{\partial t} B^{\tilde{\phi}}
= - \frac{1}{\sqrt{\gamma}} \frac{\partial}{\partial r} \left [
\alpha \sqrt{\gamma} ( E_{\tilde{\Psi}} - \sqrt{\gamma} N^\phi B^{\tilde{r}}
+ \sqrt{\gamma} N^r B^{\tilde{\phi}})
\right ]  ,  \label{faraday4b1d}\\
& 
\displaystyle \frac{\partial}{\partial t} {S}_{\tilde{r}} =
- \frac{1}{\sqrt{\gamma}} \frac{\partial}{\partial r}
[ \alpha \sqrt{\gamma} (T^{\tilde{r}}_{\tilde{r}} + {N}^r \tilde{S}_r) ] 
- \frac{\partial \alpha}{\partial r} \tilde{u}
\nonumber \\
& \displaystyle - \frac{\partial}{\partial r} (\alpha N^r) {S}_{\tilde{r}}
- \frac{\partial}{\partial r} (\alpha N^\phi) {S}_{\tilde{\phi}}
- \frac{\partial}{\partial \Psi} (\alpha N^\Psi) {S}_{\tilde{\Psi}}
+ \frac{1}{2} \frac{\partial}{\partial r} \gamma_{jk} T^{\tilde{j}\tilde{k}}
 , \\
& \displaystyle \frac{\partial}{\partial t} S_{\tilde{\phi}} =
- \frac{1}{\sqrt{\gamma}} \frac{\partial}{\partial r}
[ \alpha \sqrt{\gamma} (T^{\tilde{r}}_{\tilde{\phi}} + {N}^r S_{\tilde{\phi}}) ] , 
\label{eq2stp} \\
& \displaystyle \frac{\partial}{\partial t} e^\infty
= - \frac{1}{\sqrt{\gamma}} \frac{\partial}{\partial r} (\alpha \sqrt{\gamma} S^r) .
\label{eq2einf}
\end{eqnarray}
Noted that other equations from Eqs. (\ref{gauss2ele}), (\ref{ampere}), 
(\ref{3+1enem}), (\ref{3+1moem}), and (\ref{eq2energycons}) 
yield non--1D equations that include the $\Psi$-derivative as shown in 
Appendix \ref{non1deq}.
Eq. (\ref{gauss4b1d}) shows $\sqrt{\gamma} B^{\tilde{r}}$ is uniform and constant.
In fact, we have
$\displaystyle B^{\tilde{r}} = ^\ast F^{\tilde{0} \tilde{r}} = \frac{1}{\sqrt{\gamma}}
(\partial_\Psi A_\phi - \partial_\phi A_\Psi)= \frac{1}{\sqrt{\gamma}}$.
The poloidal component of the magnetic field $\tilde{B}_{\rm p}$ is defined by
$\tilde{B}_{\rm p} = \sqrt{\gamma_{rr}} B^{\tilde{r}} = \sqrt{\gamma_{rr}/\gamma}$.

The primitive variables $E^{\tilde{\Psi}}$ is given by
conservation quantities $B^{\tilde{r}}$, $B^{\tilde{\phi}}$,
$S_{\tilde{r}}$, $S_{\tilde{\phi}}$ as
\begin{eqnarray}
E^{\tilde{\Psi}} = \frac{1}{\sqrt{\gamma}}
\frac{S_{\tilde{r}} B^{\tilde{\phi}} - S_{\tilde{\phi}} B^{\tilde{r}}}{(\tilde{B})^2}
\label{priepsi}
\end{eqnarray}
where $B_{\tilde{r}} = \gamma_{rr} B^{\tilde{r}} + \gamma_{r\phi} B^{\tilde{\phi}}$,
$B_{\tilde{\phi}} = \gamma_{\phi r} B^{\tilde{r}} + \gamma_{\phi\phi} B^{\tilde{\phi}}$, and
$(\tilde{B})^2 = B^{\tilde{r}} B_{\tilde{r}} + B^{\tilde{\theta}} B_{\tilde{\theta}}$.
We have 
$E_{\tilde{\Psi}} = \frac{1}{\gamma^{\Psi \Psi}} (E^{\tilde{\Psi}}
- \sqrt{\gamma} N^\Psi \gamma^{\Psi r} B^{\tilde{\phi}} )$
using $E^{\tilde{\Psi}} = \gamma^{\Psi i} E_{\tilde{i}}$ with 
Eqs (\ref{cond6magsurcoo}), (\ref{cond6degeneracy}), and $\gamma^{\Psi \phi} = 0$.
We can write
\begin{eqnarray}
T^{\tilde{i}}_{\tilde{j}} = - E^{\tilde{i}} E_{\tilde{j}}
- B^{\tilde{i}} B_{\tilde{j}} + \tilde{u} \delta^i_j, \\
\tilde{u} = \frac{1}{2} ( B^{\tilde{r}} B_{\tilde{r}} 
+ B^{\tilde{\phi}} B_{\tilde{\phi}}
+(\tilde{E})^2),
\label{defutilde}
\end{eqnarray}
where $(\tilde{E})^2 = E^{\tilde{i}} E_{\tilde{i}}$.

Using Eqs. (\ref{faraday4b1d})-(\ref{eq2stp}), (\ref{priepsi})-(\ref{defutilde}) with
$B^{\tilde{r}} = B_0/\sqrt{\gamma}$ ($B_0$ is constant), 
we perform 1D FFMD numerical simulations for an arbitrary magnetic surface. 
When we use the flux coordinates, we have $B_0 =1$.


\subsection{Constants of steady state in 1D FFMD}

Here, we show the constants for the steady state in 1D FFMD.
We assume that the force-free field is stationary and axisymmetric.
First, the equation of the energy conservation (Eq. (\ref{eq2einf}))
reads the energy flux constant,
\begin{eqnarray}
P = \alpha \sqrt{\gamma} S^r.
\label{energyflux}
\end{eqnarray}
Here $e^\infty = - \chi_\nu T^{\nu 0} = \alpha (\tilde{u} + N^i S_{\tilde{i}})$, 
$S^r = - \chi_{\nu} T^{\nu r} = \frac{1}{\alpha} \epsilon^{rjk} E_j B^k
= \frac{1}{\alpha \sqrt{\gamma}} E_\Psi B_\phi$, where
$\chi^\mu = (1,0,0,0)$ is the Killing vector with respect to the time boost symmetry
and $E_{\Psi} = \alpha (E_{\tilde{\Psi}} + \epsilon_{\Psi jk} N^j B^{\tilde{k}})
= \alpha (E_{\tilde{\Psi}} + \sqrt{\gamma} N^r B^{\tilde{\phi}})$,
$B_\phi = \alpha (B_{\tilde{\phi}} - \epsilon_{\phi jk} N^j B^{\tilde{k}})
= \alpha (B_{\tilde{\phi}} + \sqrt{\gamma} N^r E^{\tilde{\Psi}})$.
We have 
\begin{eqnarray}
P = E_\Psi B_\phi = \alpha^2
(E_{\tilde{\Psi}} + \sqrt{\gamma} N^r B^{\tilde{\phi}})
(B_{\tilde{\phi}} + \sqrt{\gamma} N^r E^{\tilde{\Psi}}) .
\label{powerexpr}
\end{eqnarray}

Next, vanishing right-hand side of Eq. (\ref{faraday4b1d}) provides the constant,
\begin{eqnarray}
\Omega_{\rm F} = - \alpha (E_{\tilde{\Psi}} - \sqrt{\gamma} N^\phi B^{\tilde{r}}
+ \sqrt{\gamma} N^r B^{\tilde{\phi}}) .
\end{eqnarray}
Using the constant $\sqrt{\gamma} B^{\tilde{r}} = 1$ in the flux coordinates, we also have
\begin{eqnarray}
\Omega_{\rm F} = 
- \frac{\alpha}{\sqrt{\gamma} B^{\tilde{r}}} 
(E_{\tilde{\Psi}} - \sqrt{\gamma} N^\phi B^{\tilde{r}}
+ \sqrt{\gamma} N^r B^{\tilde{\phi}}) .
\label{omegafexp}
\end{eqnarray}
Here $\Omega_{\rm F}$ represents the angular velocity of the magnetic field line.
Last, Eq. (\ref{ampere31}) in the situation with $J^{\tilde{\Psi}} = - \tilde{\rho}_{\rm e}
N^\Psi$ yields the constant
\begin{eqnarray}
I = \alpha (B_{\tilde{\phi}} + \sqrt{\gamma} N^r E^{\tilde{\Psi}}
- \sqrt{\gamma} N^\Psi E^{\tilde{r}}) .
\label{currexp}
\end{eqnarray}
Here $I$ is the current inside of the magnetic surface of $\Psi = \Psi_0$:
$\Psi \le \Psi_0$.

Using Eqs. (\ref{omegafexp}) and (\ref{currexp}), we obtain the analytic solution of the 
steady-state force-free field with $I$ and $\Omega_{\rm F}$,
\begin{eqnarray} 
& \displaystyle B^{\tilde{\phi}} = \frac{1}{\alpha \sqrt{\gamma} D}
\left [ \sqrt{\gamma} I - \Upsilon \Omega_{\rm F} 
- \alpha (\gamma_{\phi r} + \Upsilon N^\phi) \right ] , 
\label{bphi6iomg} \\
& \displaystyle E_{\tilde{\Psi}}  = \frac{1}{\alpha D} \left [
\gamma_{\phi \phi} \Omega_{\rm F} - \sqrt{\gamma} N^r I + 
\alpha ( \gamma_{\phi \phi} N^\phi + \gamma_{\phi r} N^r ) \right ],
\label{epsi6iomg}
\end{eqnarray}
where $\Upsilon = \gamma (N^r \gamma^{\Psi \Psi} - N^\Psi \gamma^{r \Psi})$ 
and $D=\gamma_{\phi \phi} - \Upsilon N^r$.
The $I$ and $\Omega_{\rm F}$ of the steady state 
are given in Section \ref{anasol4stestafff}.
In the BL and KS coordinates, we have
$\Upsilon = 0$ and
$\Upsilon = \gamma_{\rm KS} N^r_{\rm KS} \gamma^{\theta \theta}_{\rm KS}$, respectively.

Eq. (\ref{powerexpr}) yields the relation
\begin{equation}
P = - I \Omega_{\rm F}.
\end{equation}
In the BL coordinates, these constants becomes
\begin{eqnarray}
\Omega_{\rm F} =  \omega - \frac{\alpha E_{\hat{\theta}}}{R B^{\hat{r}}},
\hspace{0.5cm}
I = \alpha R B_{\hat{\phi}}.
\end{eqnarray}
In the KS coordinates, they are
\begin{eqnarray}
\Omega_{\rm F} =  - \frac{\alpha}{\sqrt{\gamma} B^{\tilde{r}}} 
(E_{\tilde{\theta}} + \sqrt{\gamma} N^r B^{\tilde{\phi}}) ,
\hspace{0.5cm}
I = \alpha (B_{\tilde{\phi}} + \sqrt{\gamma} N^r E^{\tilde{\theta}}).
\end{eqnarray}

\subsection{Spacetime with flux coordinates for an arbitrarily given magnetic surface}


To describe the space-time around the spinning black hole with mass $M$
and angular momentum $J$, 
we use the two coordinate systems,
BL coordinates and KS coordinates, defined below.
Here we use $a = (J/J_{\rm max}) M$,
where $J_{\rm max} = M^2$ is the maximum angular momentum of a 
spinning black hole with mass $M$.
The metric of BL coordinates $x^\mu = (t, r, \theta, \phi)$
is given by
\begin{eqnarray}
ds^2 = g_{\mu\nu} dx^\mu dx^\nu = - \left ( 1 - \frac{2 M r}{\rho^2} \right ) dt^2
+ \frac{\rho^2}{\Delta} dr^2 + \rho^2 d \theta^2 + R^2 d \phi^2
- 2 \times \frac{2 M a r}{\rho^2} \sin^2 \theta d \phi dt , 
\label{boyerlindquistmet}
\end{eqnarray}
where $\rho^2 = r^2 + a^2 \cos^2 \theta$，$\Delta = r^2 -2 M r + a^2$，
$\Sigma^2 = (r^2+a^2)^2 -a^2 \Delta \sin^2 \theta$, 
$\displaystyle R^2 = \frac{\Sigma^2}{\rho^2} \sin^2 \theta$.
The horizon is given by $\Delta = 0$, which yields 
$r = M \pm \sqrt{M^2 - a^2} \equiv r_{\rm H}$.
In this coordinate system,
$\displaystyle \alpha = \sqrt{\frac{\Delta \rho^2}{\Sigma^2}}$，
$\displaystyle \beta^\phi= - \frac{2 a M r}{\Sigma^2} \equiv - \omega$,
$g = - \rho^4 \sin^2 \theta$.

The metric of KS coordinates $x^\mu = (t, r, \theta, \phi)$ is given by
\begin{eqnarray}
& \displaystyle
ds^2 = g_{\mu\nu} dx^\mu dx^\nu = - \left ( 1 - \frac{2 M r}{\rho^2} \right ) dt^2
+ \left ( 1+ \frac{2Mr}{\rho^2} \right ) dr^2
+ \rho^2 d \theta^2 + R^2 d \phi^2 \nonumber \\ 
&\displaystyle
 - 2 \times a \left ( 1 + \frac{2 M r}{\rho^2} \right ) \sin^2 \theta dr d \phi
+ 2 \times  \frac{2 M r}{\rho^2}  dt dr 
- 2 \times \frac{2 M a r}{\rho^2} \sin^2 \theta dt d \phi.
\end{eqnarray}
At the horizon $r = r_{\rm H}$,
all of the metrics of the KS coordinates are finite, while some metric of 
BL coordinates, for example, $g_{rr}$, is infinity.
In the KS coordinates, we have
$
\displaystyle  \beta^r = \frac{2 M r}{\rho^2 + 2 M r},
 \beta^\theta = 0,
 \beta^\phi = 0 $,
$\gamma = \rho^2 (\rho^2 + 2Mr) \sin^2 \theta$,$g = - \rho^4 \sin^2 \theta$.
The BL and KS coordinates are used for the base of the flux
coordinates. In the cases of radial magnetic surfaces 
as shown in Fig. \ref{ffmd1d}, we directly use these coordinates as the flux coordinates.

It is noted that BL 
$x_{\rm BL}^\mu=(t_{\rm BL}, r_{\rm BL}, \theta_{\rm BL}, \phi_{\rm BL})$
and KS
$x_{\rm KS}^\mu=(t_{\rm KS}, r_{\rm KS}, \theta_{\rm KS}, \phi_{\rm KS})$ coordinates
are related as
\begin{eqnarray}
dt_{\rm KS}  = dt_{\rm BL} + \frac{2 M r}{\Delta} dr_{\rm BL} ,   \hspace{0.5cm}
d\phi_{\rm KS}  = d\phi_{\rm BL} + \frac{a}{\Delta} dr_{\rm BL} ,   \hspace{0.5cm}
r_{\rm KS}  = r_{\rm BL},  \hspace{0.5cm}
\theta_{\rm KS} = \theta_{\rm BL} . 
\label{rel4bl2ks} 
\end{eqnarray}
Then we note that at the horizon and infinitely far from the the black hole,
the times on KS and BL
coordinates are infinitely different. That is, the finite time on
the BL coordinates corresponds to the infinite past
on the KS coordinates. Conversely, at an infinitely far point from the
black hole, the finite time on the BL coordinates corresponds
to the infinite future on the KS coordinates.

We introduce the flux coordinates $x^\mu = (t, r, \Psi, \phi)$ 
around an arbitrary axisymmetric magnetic surface $\Psi = \Psi_0$,
where $\Psi_0$ is a constant, which is described by
$\theta = \theta_0 (r)$, where $\theta_0(r)$ is function of $r$.
When the magnetic surface $\Psi = \Psi$ around the magnetic surfaces $\Psi=\Psi_0$ 
is described by $\theta = \theta(r, \Psi)$, 
the Taylor expansion of $\theta(r, \Psi)$ 
with an infinitesimally small variable $\Psi - \Psi_0$ yields
$\displaystyle \theta(r, \Psi) = \theta(r, \Psi_0) +
\left ( \frac{\partial \theta(r, \Psi)}{\partial \Psi} \right )_{\Psi=\Psi_0}
(\Psi - \Psi_0) + \cdots$. Then we have
\begin{equation}
\Psi \approx \Psi_0 + b(r) \left ( \theta - \theta_0(r) \right ),
\label{psitaylorexp}
\end{equation}
where $\displaystyle b(r) = \left [ \left  ( \frac{\partial \theta(r, \Psi)}{\partial \Psi} 
\right )_{\Psi=\Psi_0} \right ]^{-1}$ and $\theta_0(r) = \theta(r, \Psi_0)$.
The function $b(r)$ determines the flared shape of the magnetic surfaces.
The metric of the flux coordinates in the Kerr spacetime is given 
by the metric of KS coordinates $x^{\mu}_{\rm KS}$.
\begin{eqnarray}
\begin{array}{llll}
& \gamma_{r r} = \gamma_{rr}^{\rm KS} + K^2 \gamma_{\theta \theta}^{\rm KS}, &
\displaystyle \gamma_{\Psi \Psi}  = \frac{1}{b^2} \gamma_{\theta \theta}^{\rm KS}, &
\gamma_{\phi \phi}  = \gamma_{\phi \phi}^{\rm KS}, \\
& \displaystyle \gamma_{r \Psi} = \frac{K}{b} \gamma_{\theta \theta}^{\rm KS}, &
\gamma_{\Psi \phi} = 0, & 
\gamma_{r \phi} = \gamma_{r \phi}^{\rm KS}, \\
\alpha^2 = \alpha^2_{\rm KS} , & \beta^r = \beta^r_{\rm KS}, & 
\beta^\Psi = b K \beta^r_{\rm KS}, & 
\beta^\phi = 0,
\end{array}
\end{eqnarray}
where $\displaystyle K \equiv \theta_0^\prime (r) - \frac{b^\prime(r)}{b(r)^2}$. 
We also have
\begin{eqnarray}
\begin{array}{lll}
\gamma^{rr}  = \gamma^{rr}_{\rm KS} , &
\gamma^{\Psi \Psi}  = b^2 (\gamma^{\Psi \Psi}_{\rm KS} + K^2 \gamma^{rr} ), &
\gamma^{\phi \phi}  = \gamma^{\phi \phi}_{\rm KS}, \\
\gamma^{r \Psi} = K b \gamma^{rr}_{\rm KS}, &
\gamma^{\Psi \phi} = -K b \gamma^{\phi r}_{\rm KS}, &  
\gamma^{r \phi} = \gamma^{r \phi}_{\rm KS}.\\
  & & 
\end{array}
\end{eqnarray}

\section{Analytic solution of a steady-state force-free field along
an arbitrary magnetic surface
\label{anasol4stestafff}}

Here we derive the analytic solution of the steady-state force-free field along
an arbitrary magnetic surface $\Psi = \Psi_0$ using the flux coordinates 
$(t, r, \Psi, \phi)$. We derive the constants $I$ and $\Omega_{\rm F}$ 
of the steady state along the magnetic surface $\Psi = \Psi_0$,
which is indicated
by $\theta = \theta_0 (r)$. Because the constants of the steady-state force-free field
do not depend on the coordinates, we use the KS coordinates 
for the base coordinates
without loss of generality. First, we derive the condition with respect to
$I$ and $\Omega_{\rm F}$ at
the horizon, which is called the ``Znajek condition" \citep{znajek77}.
Using Eqs. (\ref{bphi6iomg}), we have
\begin{equation}
\displaystyle B^{\tilde{\phi}} = 
\frac{I + (\gamma N^r \gamma^{\Psi \Psi} \Omega_{\rm F} 
- \alpha \gamma_{\phi r} ) B^{\tilde{r}}}
{\alpha (\gamma_{\phi \phi} - (\sqrt{\gamma} N^r)^2 \gamma^{\Psi \Psi})}.
\end{equation}
Because the magnetic surface is radial at the horizon, $b$ in Eq. (\ref{psitaylorexp}) 
is constant $b_{\rm H}$ around the horizon. 
We have $\gamma^{\Psi \Psi} = b_{\rm H}^2 \gamma_{\rm KS}^{\theta \theta}$ 
and $N^\Psi = N_{\rm KS}^\theta =0$ around the horizon. We obtain
\begin{equation}
B^{\tilde{\phi}} = 
\frac{I + ({\gamma}_{\rm KS} N_{\rm KS}^r \gamma_{\rm KS}^{\theta \theta} \Omega_{\rm F} 
- \alpha_{\rm KS} \gamma^{\rm KS}_{\phi r}) B^{\tilde{r}}}
{\alpha (\gamma^{\rm KS}_{\phi \phi} 
- (\sqrt{\gamma_{\rm KS}} N_{\rm KS}^r)^2 \gamma^{\theta \theta}_{\rm KS})}.
\label{bphi6iomg2ks}
\end{equation}
Because the denominator of Eq. (\ref{bphi6iomg2ks}),
$\displaystyle \gamma^{\rm KS}_{\phi \phi} 
- (\sqrt{\gamma_{\rm KS}} N_{\rm KS}^r)^2 \gamma^{\theta \theta}_{\rm KS}
= \Delta \sin^2 \theta (1 + 2Mr/\rho^2)$ vanishes at the horizon, 
the continuity of $B^{\tilde{\phi}}$ at the horizon yields the Znajek condition,
\begin{equation}
I = \left (-\sqrt{\gamma}_{\rm KS} N_{\rm KS}^r \gamma_{\rm KS}^{\theta \theta} 
\Omega_{\rm F} + \alpha_{\rm KS}
\gamma^{\rm KS}_{\phi r} \right ) b_{\rm H}
= \frac{1}{\rho_{\rm H}^2} (2 M r_{\rm H} \Omega_{\rm H} -a ) 
b_{\rm H} \sin \theta_{\rm H}, 
\label{znajekcond}
\end{equation}
where 
$\displaystyle \rho_{\rm H}^2 = r_{\rm H}^2 + a^2 \cos^2 \theta_{\rm H}$.
Here $b_{\rm H}$ and $\theta_{\rm H}$ are the values of $b$ and $\theta$ at the horizon,
respectively.

Next, we derive the condition with respect to $I$ and $\Omega_{\rm F}$ 
at infinity. The condition is given by the condition of outward propagation 
of the force-free electromagnetic wave, 
\begin{equation}
E_{\hat{\Psi}} = B^{\hat{\phi}}
\label{cond6mickel}
\end{equation}
where $\wedge$ indicates the orthonormal (proper) frame of the normal observer, 
in which the $r$-coordinate is parallel to the magnetic surface $\Psi = \Psi_0$.
At infinity, the proper frame $x^{\hat{\mu}}$ is given by 
$d \hat{t} = d \tilde{t}$,
$d \hat{r} = h_1 (d \tilde{r} + k d \tilde{\Psi})$, 
$d \hat{\Psi} = h_2^\prime d \tilde{\Psi}$, 
and $d \hat{\phi} = h_3 (d \tilde{\phi} + l d \tilde{r})$ where $h_1^2 = \gamma_{rr} = 1 + K^2 r^2 - h_3^2 l^2$,
$\displaystyle h_1^2 k = \frac{K}{b} r^2$, $\displaystyle h_2^2 
= \gamma_{\theta \theta} = \frac{r^2}{b}$,
$h_3^2 = \gamma_{\phi \phi} =  r^2 \sin^2 \theta$,
$h_3^2 l = -a \sin^2 \theta$, and $(h_2^\prime)^2 = h_2^2 - h_1^2 k^2$ at infinity.
Using the proper frame at infinity, we have 
$ds^2 = -d \hat{t}^2 +d \hat{r}^2+d \hat{\Psi}^2+d \hat{\phi}^2$.
At infinity, we have $h_1^2 = 1 - K^2 r^2$, $\displaystyle k = \frac{K}{b}
\frac{r^2}{1- K^2 r^2}$, and $l=0$.
The relationships between the variables of the proper frame and the normal observer frame
\begin{equation}
E_{\hat{\Psi}} = \frac{1}{h'_2} (E_{\tilde{\Psi}} - k E_{\tilde{r}}),
B^{\hat{\phi}} = h_3 B^{\tilde{\phi}},
\end{equation}
and Eq. (\ref{cond6mickel}) yield 
\begin{equation}
I = - b_\infty \sin \theta_\infty \Omega_{\rm F}
\label{ieqqsinqomgf}
\end{equation}
at infinity, where 
$\displaystyle b_\infty \sin \theta_\infty = \lim_{r \to \infty} b \sin \theta $.
Using Eqs. (\ref{znajekcond}) and (\ref{ieqqsinqomgf}), we obtain the 
constant
\begin{equation}
\Omega_{\rm F} 
= \frac{a}{2 M r_{\rm H} + \rho^2_{\rm H} 
b_\infty \sin \theta_\infty/b_{\rm H} \sin \theta_{\rm H}} .
\label{omgfdtmd}
\end{equation}
The constant $I$ is given by (\ref{ieqqsinqomgf}) and (\ref{omgfdtmd}) as
\begin{equation}
I = - \frac{a b_\infty \sin \theta_\infty}
{2 M r_{\rm H} + \rho^2_{\rm H} b_\infty \sin \theta_\infty/b_{\rm H} \sin \theta_{\rm H}}.
\label{idtmd}
\end{equation}
Eqs. (\ref{omgfdtmd}) and (\ref{idtmd}) are the generalization of the solution
given by \citet{menon05}.
The electric field $E_{\tilde{\Psi}}$ and the magnetic field $B^{\tilde{r}}$ of the steady-state solution
are given by Eqs. (\ref{bphi6iomg}) and (\ref{epsi6iomg}) 
with $\Omega_{\rm F}$ and $I$ given by
Eqs. (\ref{omgfdtmd}) and (\ref{idtmd}), respectively,
for the flux coordinates on the BL and KS coordinates.

In general, when the magnetic surface is radial ($b_\infty = b_{\rm H}$, 
$0< \theta_\infty = \theta_{\rm H}< \pi$), we have
\begin{equation}
\Omega_{\rm H} = \frac{a}{(r_{\rm H} + 2 M)r_{\rm H}}
= \frac{a}{4 M r_{\rm H} - a^2}.
\label{omegah4rad}
\end{equation}
In the case of $a \ll M$, Eq. (\ref{omegah4rad}) yields the well-known relation
$\displaystyle \Omega_{\rm F} = \frac{\omega_{\rm H}}{2}$
where $\displaystyle \omega_{\rm H} = -\frac{g_{\phi t}}{g_{\phi \phi}}
= \frac{a}{2 M r_{\rm H}}$ is the angular velocity of the normal observer
at the horizon.

\subsection{Derivation of two Blandford--Znajek solutions with
perturbation method}

To show the validity of the 1D FFMD analytic solution, we derive the analytic
solutions of the steady-state force-free field around a extremely slowly
spinning black hole ($a \ll M$) given by \citet{blandford77}. 
\citet{blandford77} resorted to a perturbation method in which
they expanded on the powers of $a/M$. Such a technique can only be of use
when the change in the poloidal field caused by spinning up a non-rotating
field configuration (supported by currents in an equatorial disk) can be
regarded as small. They wrote an exact axisymmetric vacuum solution for the 
magnetic field in a Schwarzschild metric by $\Psi (r, \theta) = X(r, \theta)$
used as the unperturbed function.
According to the perturbation method,
the perturbed variables, the electromagnetic angular frequency and
the current $I$ are written by $\displaystyle \Omega_{\rm F}
= \frac{a}{M^2} W(r, \theta)$ and
$\displaystyle I = \frac{a}{M^2} Y(r, \theta)$, respectively.
They showed two examples of the perturbation technique:
(a) the radial magnetic field (of opposite polarity in the two hemisphere)
and (b) a force-free magnetosphere in which the magnetic field lines lie
on paraboloidal surfaces (cutting an equatorial disk).
We derive the expression of $W$ and $Y$ using the magnetic surfaces
for both cases as follows.

\subsubsection{The case of the radial magnetic surface}

The vector potential
\begin{equation}
\Psi(r, \theta) = X(r, \theta) = - C \cos \theta \hspace{1cm}
(0 \le \theta \le \pi/2)
\label{monopole0}
\end{equation}
with a constant $C$ is an exact solution of the vacuum Maxwell equations
in a Schwarzschild metric, which describes the unperturbed radial magnetic field
(Eq. (6.1) in \citet{blandford77}).
\citet{blandford77} used this solution as the unperturbed vector potential $\Psi$ of
the split monopole field around a extremely slowly spinning black hole.
It is noted that in the case of the split monopole, the solution is used
except on an equatorial disk containing a toroidal surface current density
$I_{\rm sm} = 2C/r^2$. Eq. (\ref{monopole0}) yields $b = [\partial \theta(r, \Psi)/
\partial \Psi]^{-1} = C \sin \theta$. Then we have 
$b_{\rm H} \sin \theta_{\rm H} = b_{\infty} \sin \theta_{\infty}$ 
because $\Psi = - C \cos \theta_{\rm H} = - C \cos \theta_{\infty}$ 
reads $\theta_{\rm H} = \theta_\infty$.
Eq. (\ref{omgfdtmd}) yields 
$\displaystyle \Omega_{\rm F} = \frac{a}{4 M r_{\rm H} - a^2 \cos^2 \theta_{\rm H}}$.
Using $a \ll M$, we have $\displaystyle \Omega_{\rm F} = \frac{a}{8 M^2} = \frac{a}{M}W$.
Eq. (\ref{idtmd}) yields $\displaystyle I = - \frac{a C}{8 M^2} \sin^2 \theta
= \frac{a}{M^2} Y$. 
Then we obtain
\begin{equation}
W = \frac{1}{8},   Y = - \frac{C}{8} \sin^2 \theta.
\end{equation}
These $W$ and $Y$ are the same expressions given by Eq. (6.5) in \cite{blandford77}
in the radial magnetic surface case.

The power per solid angle from the horizon is given by
\begin{equation}
P_{\rm BZ} = \frac{d L}{d \Omega} = \frac{2 \pi d \Psi}{d \Omega} \frac{d L}{2 \pi d \Psi}
= \frac{2 \pi d \Psi}{2 \pi \sin \theta d \theta} P 
= \frac{b_{\rm H}}{\sin \theta_{\rm H}} P,
\label{powerbz}
\end{equation}
where $L$ is the power radiated by the black hole in the region between
the pole and the magnetic surface $\Psi$ and $\Omega$ is the solid angle.
We have 
\begin{equation}
P_{\rm BZ} = \left ( \frac{a^2 C}{8 M^2} \right )^2 \sin^2 \theta.
\end{equation}

\subsubsection{The case of the paraboloidal magnetic surface}

The vector potential of the paraboloidal magnetic surface is given by 
Eq. (7.1) in \cite{blandford77}
as
\begin{equation}
\Psi(r, \theta) = X(r, \theta) 
= \frac{1}{2} [ r (1 - \cos \theta) 
+ 2 M ( 1 + \cos \theta) (1 - \ln ( 1 + \cos \theta)) ] .
\label{paraboloidal0}
\end{equation}
Eq. (\ref{paraboloidal0}) yields $\displaystyle b = [\partial \theta(r, \Psi)
/\partial \Psi]^{-1}
= \frac{C}{2} \sin \theta [r + 2M \ln ( 1 + \cos \theta) ]$. 
With $a \ll M$, we have
\[
\Psi_{\rm H} = \frac{C}{2} [ 2M(1-\cos \theta_{\rm H}) 
+ 2M (1+\cos \theta_{\rm H}) (1-\ln (1+\cos \theta_{\rm H})) ] , 
\Psi_\infty = \frac{C}{2} \left [ \frac{1}{2} r_\infty \theta^2_\infty + 4M (1 - \ln 2)
\right ].
\]
Using $\Psi_{\rm H} = \Psi_\infty$, we get 
$\displaystyle \frac{1}{2} r_\infty \theta^2_\infty = 2M[
2 \ln 2 - (1+\cos \theta_{\rm H}) \ln (1+\cos \theta_{\rm H})]$.
We calculate as
\[
b_{\rm H} \sin \theta_{\rm H} = CM \sin^2 \theta_{\rm H} [1+\ln (1+\cos \theta_{\rm H})],
b_{\infty} \sin \theta_{\infty} \approx C r_\infty (1-\cos \theta_{\infty})
\approx C \frac{1}{2} r_\infty \theta^2_\infty .
\]
Using Eqs. (\ref{omgfdtmd}) and (\ref{idtmd}), we obtain 
\[
\Omega_{\rm F} = \frac{a}{M^2} \frac{N}{D},
I = -  \frac{a}{M^2} \frac{N}{D} b_\infty \sin \theta_\infty,
\]
where 
$N=\displaystyle \frac{1}{4} \sin^2 \theta_{\rm H} [1 + \ln (1+\cos \theta_{\rm H})]$ and 
$D = 4 \ln 2 + \sin^2 \theta_{\rm H} +\{ \sin^2 \theta_{\rm H} -2 (1+\cos \theta_{\rm H}) \}
\ln (1+\cos \theta_{\rm H})$.
Then we have 
\begin{equation}
W = \frac{N}{D},
Y = 2MC [2 \ln 2 - (1+ \cos \theta_{\rm H}) \ln (1+ \cos \theta_{\rm H})] \frac{N}{D} .
\end{equation}
The expression $W$ is identified to the perturbed solution
given by Eq. (7.5) in \citet{blandford77}.

Using Eq. (\ref{powerbz}), we obtain the power per unit solid angle,
\begin{equation}
P_{\rm BZ} = 2 \left (\frac{a}{M} \right )^2 C^2 
\{ 1 + \ln (1 + \cos \theta_{\rm H}) \}
\{ 2 \ln 2 - (1 + \cos \theta_{\rm H}) \ln (1 + \cos \theta_{\rm H}) \}
\left ( \frac{N}{D} \right )^2 .
\end{equation}

\section{Test numerical calculations of 1D FFMD on the 
magnetic surface around the equatorial plane 
\label{sec4}}

\subsection{1D FFMD equations and flux coordinates 
for the magnetic surface along the equatorial plane for numerical simulations}

We show the 1D FFMD equations for different types magnetic surface along 
the equatorial plane for numerical calculations. 
In the KS coordinates, using $x^\mu = (t, r, \theta, \phi) = (t, r, \Psi, \phi)$, 
we have the Maxwell equations and the conservation law of momentum
at the equatorial plane as follows:
\begin{eqnarray}
& \displaystyle  \frac{\partial}{\partial t} B^{\tilde{\phi}}
= - \frac{1}{\sqrt{\gamma}} \frac{\partial}{\partial r} \left [
\alpha \sqrt{\gamma} ( E_{\tilde{\theta}} 
+ \sqrt{\gamma} N^r B^{\tilde{\phi}})
\right ] , \label{faraday4bp2ks} \\
 & \displaystyle  
 \frac{\partial}{\partial t} {S}_{\tilde{r}} =
 - \frac{1}{\sqrt{\gamma}} \frac{\partial}{\partial r}
[ \alpha \sqrt{\gamma} (T^{\tilde{r}}_{\tilde{r}} + {N}^r S_{\tilde{r}}) ] 
- \frac{\partial \alpha}{\partial r} \tilde{u}
\nonumber \\
& \displaystyle  - \frac{\partial}{\partial r} (\alpha N^r) S_{\tilde{r}}
- \frac{1}{2} \frac{\partial}{\partial r} \gamma_{jk} T^{\tilde{j}\tilde{k}} , 
\label{enecon4sr2ks} \\
& \displaystyle  
\frac{\partial}{\partial t} S_{\tilde{\phi}} =
- \frac{1}{\sqrt{\gamma}} \frac{\partial}{\partial r}
[ \alpha \sqrt{\gamma} (T^{\tilde{r}}_{\tilde{\phi}} + {N}^r S_{\tilde{\phi)}} ] .
\label{enecon4sp2ks}
\end{eqnarray}
Here we use $N^\theta = 0$ and $N^\phi = 0$ at the equatorial plane 
in the KS coordinates.
%

In the BL coordinates, we have $N^r=0$, $N^\theta = 0$, 
$\alpha N^\phi = \omega$
and $g_{ij} = h_i^2 \delta_{ij}$. We use the ZAMO frame $x^{\hat{i}}$
easily and obtain the 3+1 formalism equation with the ZAMO frame
along the equatorial plane as follows:
\begin{eqnarray}
& \displaystyle  \frac{\partial}{\partial t} B^{\hat{\phi}}
= - \frac{1}{h_r h_\theta} \frac{\partial}{\partial r} \left [
h_\theta (\alpha E_{\hat{\theta}} - h_\phi \omega B^{\hat{r}})
\right ] , 
\label{faraday4bp2zamo} \\
& \displaystyle  
\frac{\partial}{\partial t} S_{\hat{r}} =
- \frac{1}{\alpha^2 \sqrt{\gamma}} \frac{\partial}{\partial r}
\left [ \frac{\alpha^2 \sqrt{\gamma}}{h_r} T^{\hat{r}}_{\hat{r}} \right ] 
- \frac{1}{h_r} \frac{\partial \alpha}{\partial r} (\hat{u} - T^{\hat{r}}_{\hat{r}})
\nonumber \\
& \displaystyle  - \sum_{i \ne r} \frac{\alpha}{h_r h_j} \frac{\partial h_i}{\partial r}
T^{\hat{j}\hat{j}} - \sigma S_{\hat{\phi}}
, \\
& \displaystyle  
\frac{\partial}{\partial t} {S}_{\hat{\phi}} =
- \frac{1}{h_\phi \sqrt{\gamma}} \frac{\partial}{\partial r}
\left [ \frac{\alpha \sqrt{\gamma} h_\phi}{h_r}
T^{\hat{r}}_{\hat{\phi}} \right ] ,
\label{enecon4sp2zamo}
\end{eqnarray}
where $\sigma = (h_\phi/h_r) \partial (\alpha N^\phi)/\partial r$.
%
For the numerical calculation of 1D FFMD, 
we use Eqs. (\ref{faraday4bp2ks})--(\ref{enecon4sp2ks}) and 
Eqs. (\ref{faraday4bp2zamo})--(\ref{enecon4sp2zamo})
for the numerical calculations in the KS and BL coordinates, respectively.

\begin{figure} 
\begin{center}
\includegraphics[width=10cm]{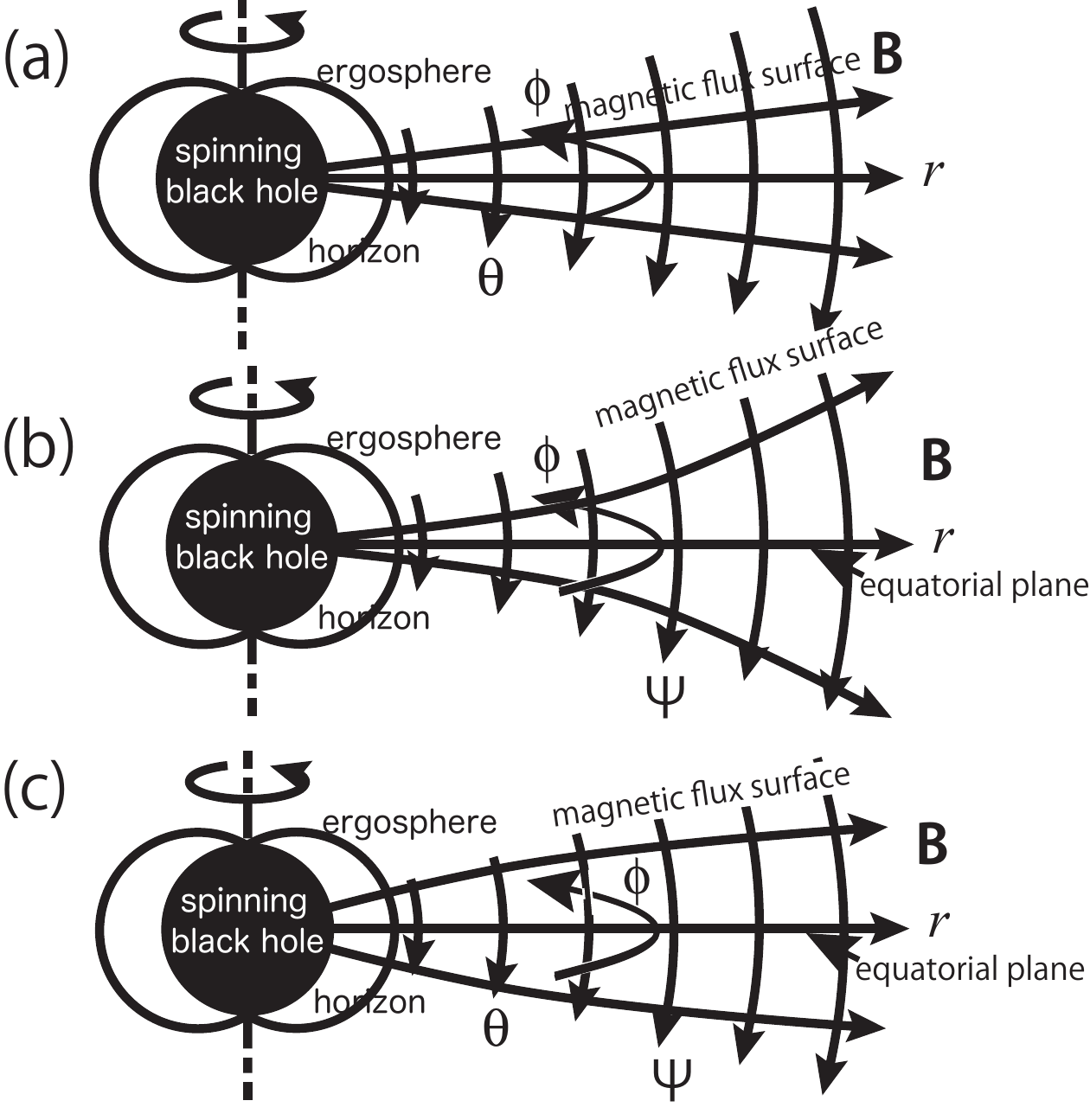}
\caption{Flux coordinates for 1D FFMD
on three types of magnetic surfaces around the equatorial plane:
(a) radial magnetic surface ($m=0$), (b) excurvature-flared magnetic surface ($m<0$),
(c) incurvature-flared magnetic surface ($m>0$). (Modified from \citet{imamura19})
\label{variousmagsurf}}
\end{center}
\end{figure}

The flux coordinates for the three types of the magnetic surface 
along the equatorial plane as shown in Fig. \ref{variousmagsurf} are given as follows:
\begin{eqnarray}
& \displaystyle  \theta_0(r)  =\frac{\pi}{2}, \nonumber \\
b(r) & =  
\left \{ 
\begin{array}{cc}
[ 1 + 0.25 (r-r_{\rm H})^2]^m & (r > r_{\rm H}) \\
1 & (r \le r_{\rm H}) \end{array}
\right . .
\label{eq4q}
\end{eqnarray}
The incurvature-flared, radial, and excurvature-flared magnetic surfaces
are given by Eq. (\ref{eq4q}) with $m > 0$, $m=0$, and $m < 0$, respectively.

\subsection{Numerical method}

We use the Lax--Wendroff scheme and the simplified TVD scheme
\citep{davis84}
for the 1D FFMD numerical calculations.
The mesh number of the standard case is 1,000 (and 10,000 if necessary).
In the KS coordinates, we use the even interval mesh
with the range $r_{\rm min} \le r \le r_{\rm max}$
($r_{\rm min} = 0.9 r_{\rm H}$, $r_{\rm max} = (40 \sim 50) r_{\rm H}$).

In the BL coordinates, we use the  even interval mesh
on the (pseudo-)tortoise coordinate for the radial coordinate $r_\ast$
such as $\displaystyle d r_\ast \propto \frac{1}{r - r_0} dr$
($r_0 < r_{\rm min}$ is a constant, and typically we set $r_0 = 1.9999$
for the $a = 0.01 M$ case).
When $r_0 = r_{\rm H}$, the coordinate $r_\ast$ becomes the exact tortoise coordinate.
We use the the equations of the Faraday law and the conservation equation
of energy flux $S_{\tilde{i}}$ ($i=r, \phi$).

\subsection{Test numerical calculations of 1D FFMD along the equatorial plane}

Here we show some test calculations of 1D FFMD simulations for variously shaped 
magnetic surfaces
around the equatorial plane, as shown in Fig. \ref{variousmagsurf}. 
In the case of the radial magnetic surface, we can use the BL
or KS coordinates as the flux coordinates.
The 1D FFMD calculations along the equatorial plane are listed in Table \ref{tabl4tests}.

\begin{table}[h]
\begin{tabular}{|l|c|c|l|l|c|c|} \hline
Magnetic surface & $a_\ast$ & \parbox[c]{2cm}{Coordinate system} 
& Initial condition & Result & Fig. No. & Sections \\ \hline
Radial & 0.01 & BL & BZ solution & Stationary & \ref{bzblk} & \ref{rad01stat} \\ \cline{3-6}
       &      & KS & BZ solution & Stationary & \ref{bzksc} & \\ \cline{3-7}
       &      & BL & Vacuum & Stationary & \ref{vacblk} & \ref{rad01vac} \\ \cline{3-7}
       &      & BL & $I = 0$ & Out/inward tsunami & \ref{kerr3tvdk}, \ref{kerr3tvdkt8} &
\ref{rad01tsun} \\ \cline{3-6}
       &      & KS & $I = 0$ & Outward tsunami & \ref{bzd1} & \\ \cline{2-7}
       & 0.95 & BL & Static solution & Stationary & \ref{bzblk_a95} & \ref{rad95stat} \\ \cline{3-6}
       &      & KS & Static solution & Stationary & \ref{finpowsol} &  \\ \cline{3-7}
       &      & KS & Pulse at $r=10M$ & Out/inward pulse propagation & \ref{fipoads_far} 
& \ref{rad95puls} \\ \cline{3-6}
       &      & KS & Pulse in ergosphere & Out/inward pulse propagation & \ref{fipoads_ergo} & \\ \cline{3-6}
       &      & KS & Pulse in horizon & Inward pulse propagation & \ref{fipoads_ihz} & \\ \cline{3-7}
       &      & BL & $I = 0$, $\Omega_{\rm F} = 0$ & Out/inward tsunami & \ref{kerr3tvdk_a95t40},\ref{kerr3tvdk_a95nea} & \ref{rad95tsun} \\ \cline{3-6}
       &      & KS & $I = 0$, $\Omega_{\rm F} = 0$ & Outward tsunami & \ref{allzero60} & \\ \hline
Incurvature-flared & 0.95 & \parbox[c]{2cm}{Flux coordinates} 
& Steady state & Steady & \ref{parabostat} & \ref{par95} \\ \cline{4-6}
       &         & based on KS  & $I=0$, $\Omega_{\rm F} = 0$ 
& Slow conversion to steady state & \ref{parabo} & \\ \hline
Excurvature-flared & 0.95 & \parbox[c]{2cm}{Flux coordinates} 
& Steady state & Unstable & --- & \ref{qua95} \\ \cline{4-6}
       &         & based on KS & $I=0$, $\Omega_{\rm F} = 0$ & No conversion & \ref{quadrapole} & 
\\ \hline
\end{tabular}
\caption{1D FFMD test calculations for magnetic surfaces at the equatorial plane
(BL=Boyer--Lindquist coordinates; KS=Kerr--Schild coordinates; 
BZ=Blandford--Znajek) \label{tabl4tests}}
\end{table}

\subsubsection{Test calculations of the Blandford--Znajek solution \label{rad01stat}}

First we calculate the Blandford--Znajek solution on the monopole
radial magnetic field as a test calculation of our numerical code
in the BL and KS coordinates.
We use the Blandford--Znajek solutions for the initial condition
of time-development numerical simulations to check the 1D FFMD code
due to the solutions being stationary.
Note that we assume that the spin parameter of black hole $a_\ast \equiv a/M$ is much
smaller than unity ($a_\ast =0.01$).
In the BL coordinates, the solution is given by
\begin{eqnarray}
& \displaystyle  B^{\hat{r}} = \frac{B_0}{h_\theta R} \left [ 
1 - a_\ast^2 f(r) \right ] \approx \frac{B_0}{h_\theta R} , 
 \hspace{1cm} B^{\hat{\theta}} = 0 ,
& \displaystyle  B^{\hat{\phi}} = - \frac{a B_0}{8 \alpha R M^2} , 
\label{bzsol4br} \\
& \displaystyle  E_{\hat{\theta}} = - \frac{R}{\alpha} (\Omega_{\rm F} - \omega) B^{\hat{r}} , 
\hspace{1cm} E_{\hat{r}} = E_{\hat{\phi}} = 0 , &
\label{bzsol4erep}
\end{eqnarray}
where $\Omega_{\rm F} = a/(8M^2)$ and $f(r)$ is a complex
function, which is given in Section 6 of Blandford \& Znajek (1977). 
Here we use the flux coordinates 
$(t, r, \Psi, \phi) = (t, r, \theta, \phi)$; thus we must set $B_0=1$.
In the Blandford--Znajek solution, the slowly spinning black hole limit condition
($a_\ast \ll 1$)
is used and we neglect the term of $f(r)$.

In the KS coordinates, the Blandford--Znajek solution is given by
\begin{eqnarray}
& \displaystyle  B^{\tilde{r}} = \frac{B_0}{\sqrt{\gamma}} ,
\hspace{1cm} B^{\tilde{\theta}} = 0 , 
& \displaystyle  B^{\tilde{\phi}} = - \frac{a \alpha B_0}{2 M r^3} \left (
1 + \frac{r}{4M} \right ) ,
\label{bzsol4br2ks} \\
& \displaystyle  E_{\tilde{\theta}} = - \frac{a B_0}{8 M^2 \alpha} - \sqrt{\gamma} 
N^r B^{\tilde{\phi}} , 
& \displaystyle  E_{\tilde{r}} = E_{\hat{\phi}} = 0 .
\label{bzsol4erep2ks}
\end{eqnarray}

\begin{figure} 
\includegraphics[width=16cm]{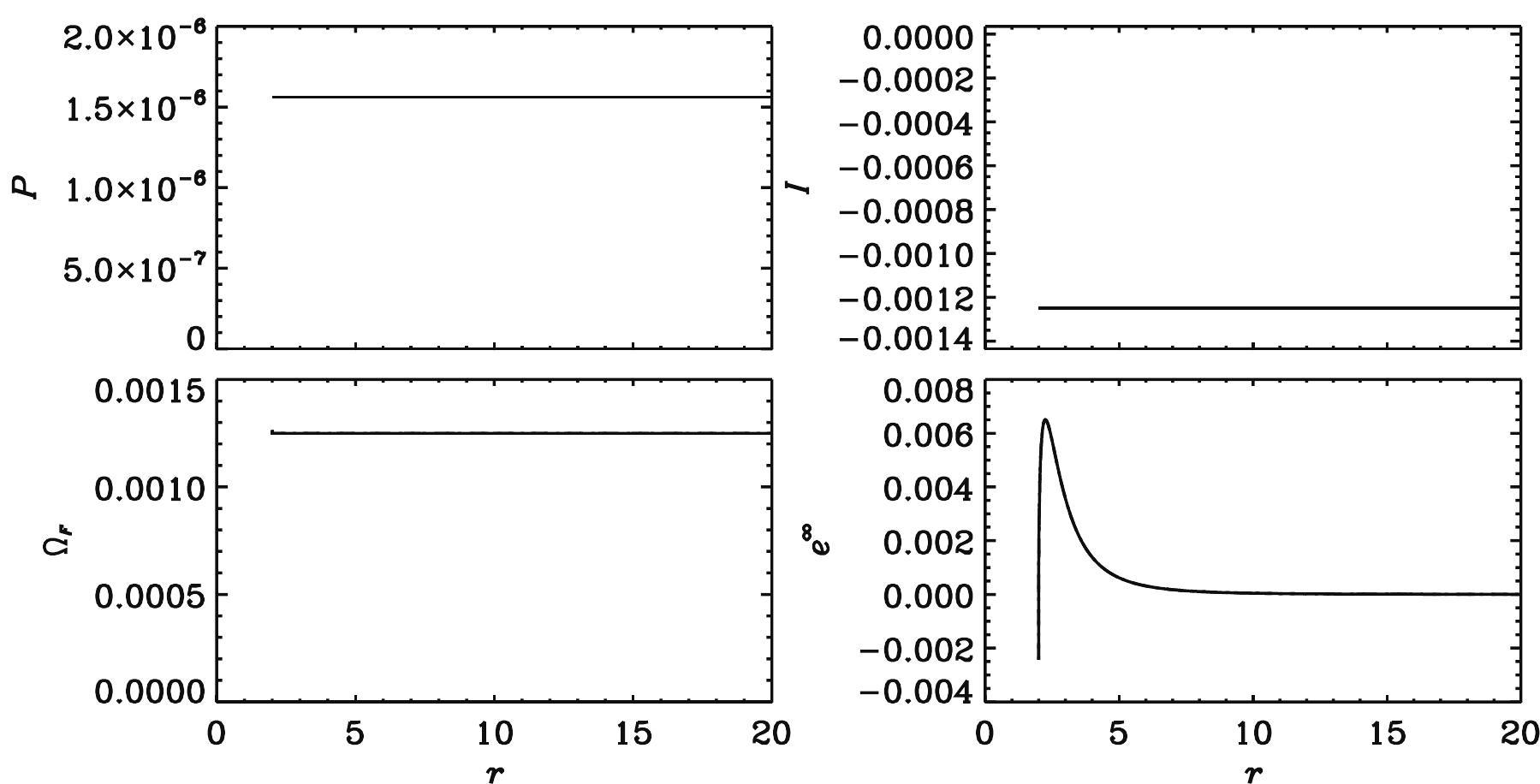}
\caption{Test by the Blandford--Znajek solution ($a_\ast=0.01$) with the BL
coordinates. Dashed lines show quantities at $t=0$,
dotted lines show results at $t=5M$, and
solid lines show results at $t=10M$.
Dashed, dotted, and solid lines overlap in this case.
\label{bzblk}}
\end{figure}

\begin{figure}[h!]
\includegraphics[width=16cm]{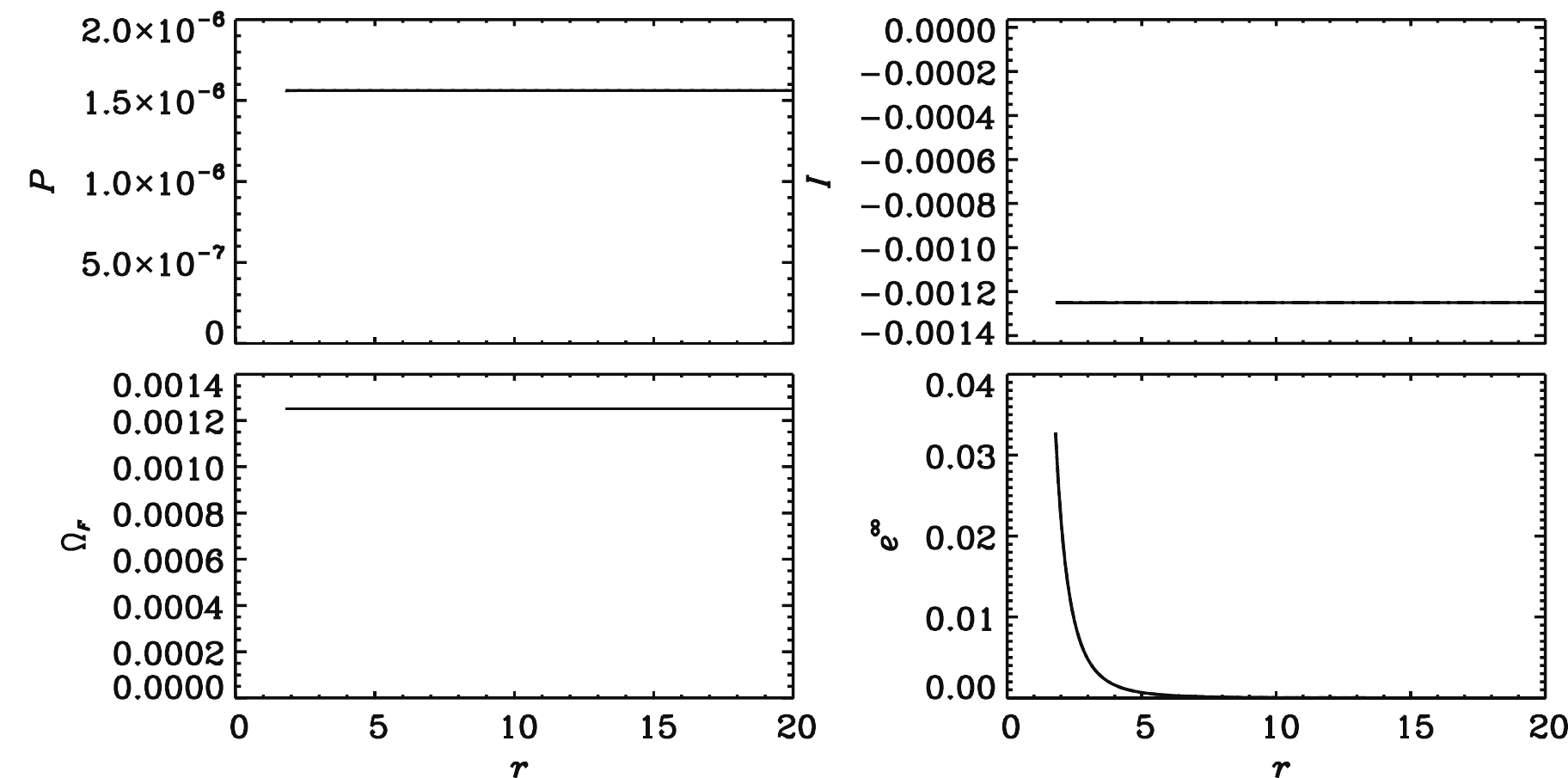}
\caption{Test by the Blandford--Znajek solution ($a_\ast=0.01$) with the KS
coordinates. Dashed lines show quantities at $t=0$,
dotted lines show results at $t=5M$, and
solid lines show results at $t=10M$.
Dashed, dotted, and solid lines overlap in this case.
\label{bzksc}}
\end{figure}

In both the BL and KS coordinates, 
the numerical results
demonstrate that the solutions are stationary,
as shown in Figs. \ref{bzblk} and \ref{bzksc}
, respectively.

\subsubsection{Test calculation of the electromagnetic field in a vacuum \label{rad01vac}}

In the force-free condition, we have the solution of the electromagnetic field
in a vacuum. It is obtained from the Maxwell equations with $\rho_{\rm e}=0$ and 
$\VEC{J} = \VEC{0}$. In the BL coordinates, we have
\begin{eqnarray}
& \displaystyle  B^{\hat{r}} = \frac{B_0}{h_\theta R} ,
\hspace{1cm} B^{\hat{\theta}} = 0 , 
& \displaystyle  B^{\hat{\phi}} = 0 , \\
& \displaystyle  E_{\hat{\theta}} = h_\theta N^\phi B^{\hat{r}} 
= \frac{R \omega}{\alpha} B^{\hat{r}} , 
& \displaystyle  E_{\hat{r}} = E_{\hat{\phi}} = 0
\end{eqnarray}

In the KS coordinates, the solution of the electromagnetic field
in a vacuum is given by 
\begin{eqnarray}
& \displaystyle  B^{\tilde{r}} = \frac{B_0}{\sqrt{\gamma}} ,
\hspace{1cm} B^{\tilde{\theta}} = 0 , 
& \displaystyle  B^{\tilde{\phi}} = - \frac{a B_{\tilde{r}}}{\Delta (1 - 2 M/r)r^2} , \\
& \displaystyle  E_{\tilde{\theta}} = -  \sqrt{\gamma} B^{\tilde{r}} B^{\tilde{\phi}} , 
& \displaystyle  E_{\tilde{r}} = E_{\hat{\phi}} = 0 .
\end{eqnarray}
Note that $B^{\tilde{\phi}}$ diverges at the horizon in the Kerr--Schild coordinates,
whose metrics are all finite. This indicates that divergence of the solution is 
physical, and we cannot use this solution.

\begin{figure}[h!]
\includegraphics[width=16cm]{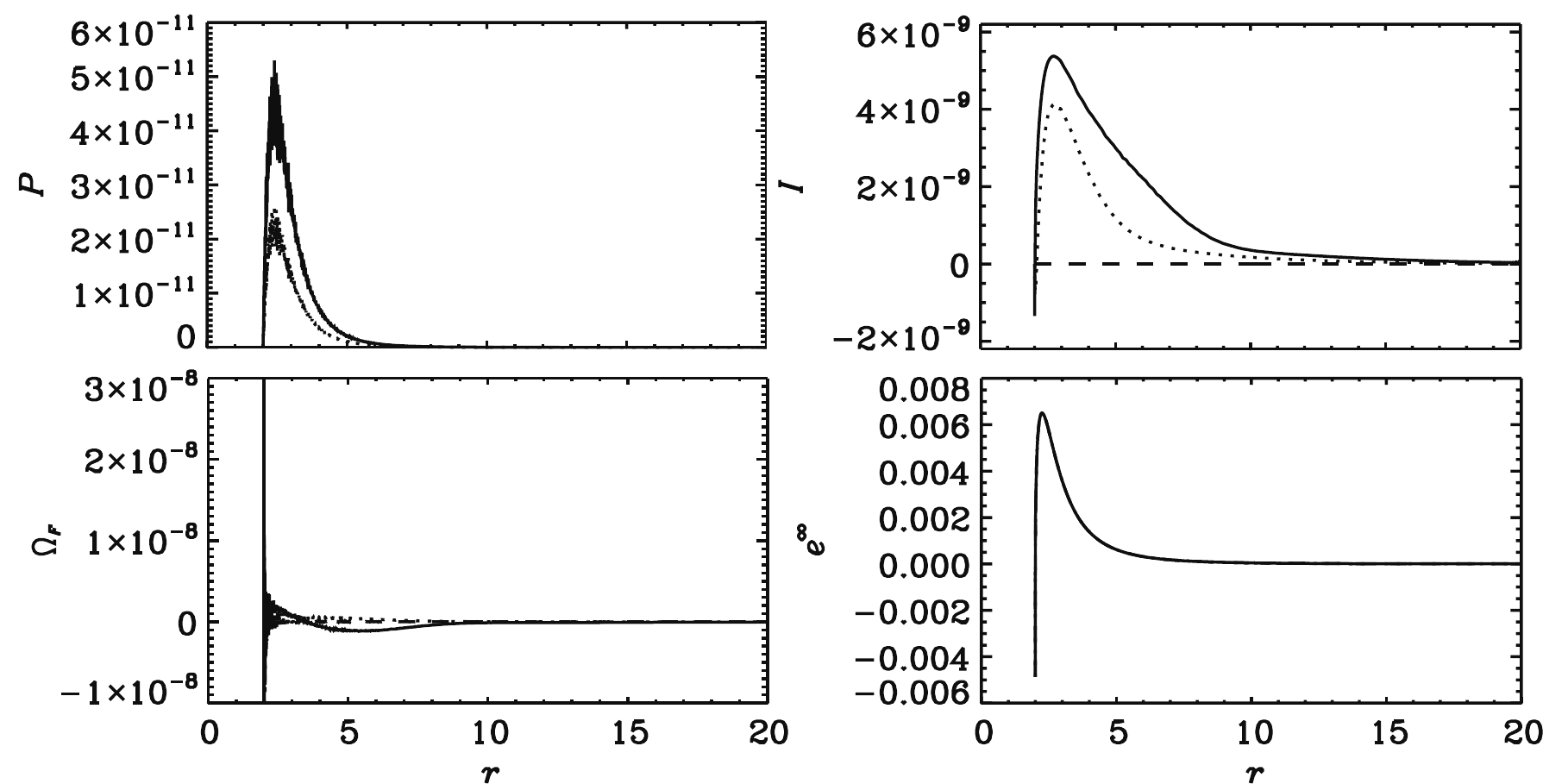}
\caption{Test by vacuum solution for a radial stationary
magnetic field  with the BL
coordinates ($a_\ast=0.01$). Dashed lines show quantities at $t=0$,
dotted lines show results at $t=5M$, and
solid lines show results at $t=10M$.
Here, $P$, $I$, and $\Omega_{\rm F}$ are negligibly small.
\label{vacblk}}
\end{figure}

We perform a test calculation with the vacuum solution in the BL coordinates,
as shown in Fig. \ref{vacblk}.
The result shows that $I$, $P$, and $\Omega_{\rm F}$ are negligibly small, and, as expected,
the solution is nearly stationary.
The numerical calculations in Subsections \ref{rad01stat} and \ref{rad01vac} indicate
the reliability of the 1D FFMD code.

\subsubsection{Test calculations of 
the dynamic process toward the Blandford--Znajek solution
with small $a_\ast$
\label{rad01tsun}}

\begin{figure}[h!]
\includegraphics[width=16cm]{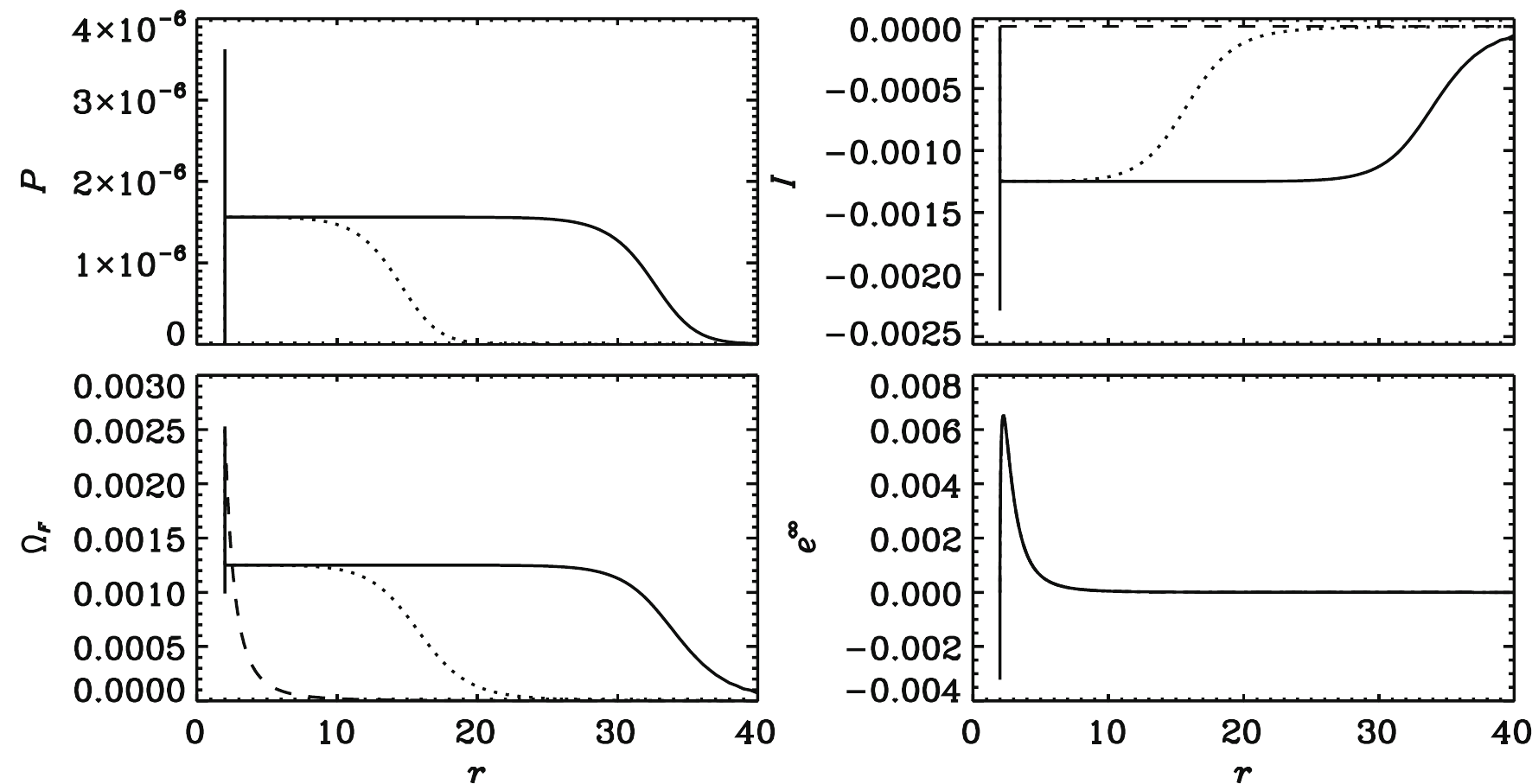}
\caption{
Simulation of FFMD with zero power and current
($P=0$, $I=0$) as initial conditions along the radial magnetic field
at the equatorial plane in terms of the BL coordinates 
($a_\ast=0.01$). 
Dashed lines show quantities at $t=0$,
dotted lines show results at $t=20M$, and
solid lines show results at $t=40M$.
\label{kerr3tvdk}}
\end{figure}

\begin{figure}[h!]
\includegraphics[width=16cm]{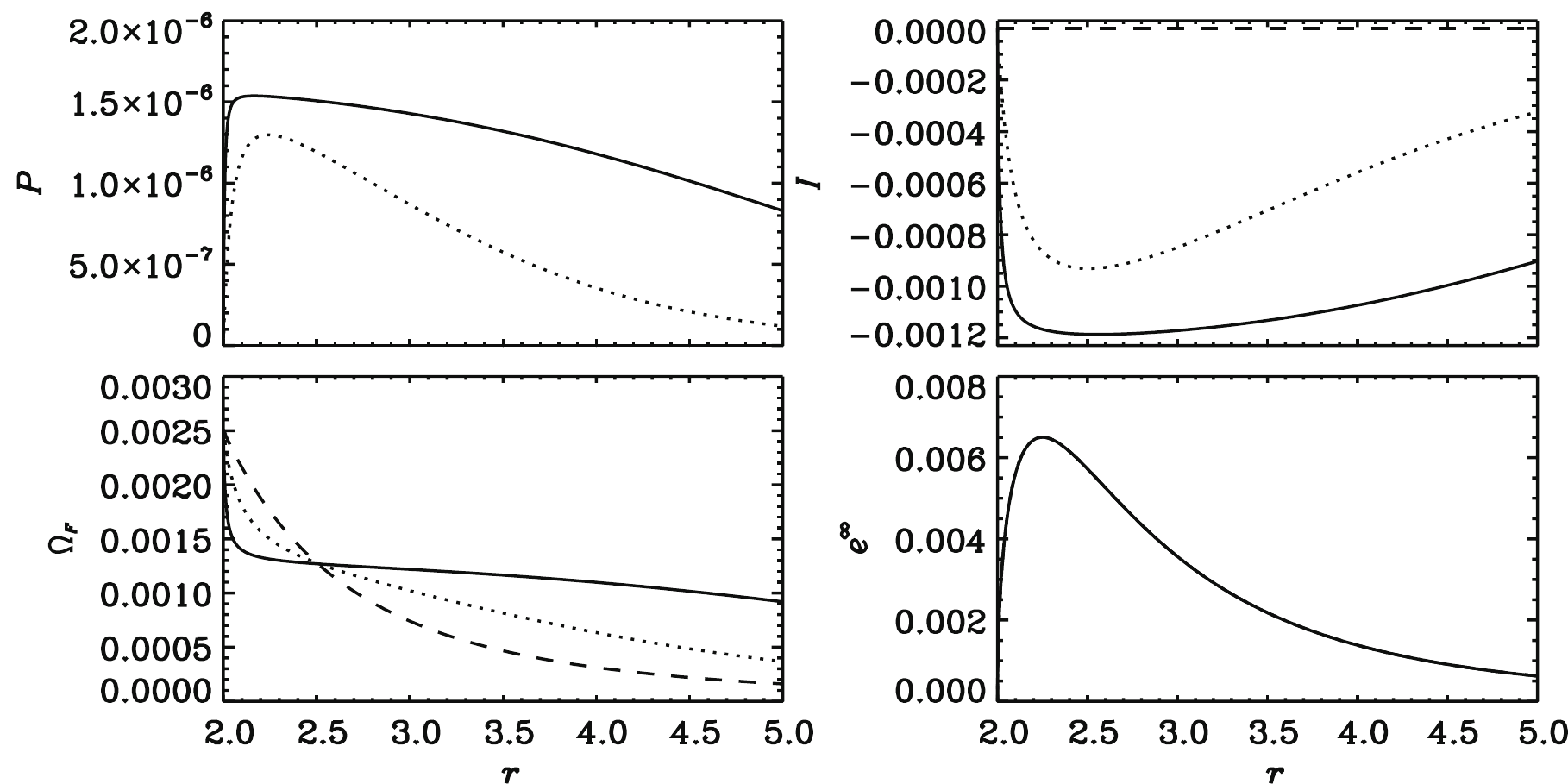}
\caption{
Enlarged plot near the horizon at an early stage of the simulation 
(initial conditions: $P=0$, $I=0$ along the radial magnetic surface at the
equatorial plane) of Fig. \ref{kerr3tvdk}
in terms of the BL coordinates ($a_\ast=0.01$). 
Dashed lines show quantities at $t=0$,
dotted lines show results at $t=4M$, and
solid lines show results at $t=8M$.
\label{kerr3tvdkt8}}
\end{figure}

Here we perform numerical test calculations for the emergence of the Blandford--Znajek
mechanism around a slowly spinning black hole of $a_\ast=0.01$
with both BL and KS coordinates.
Note that the initial condition $I=0$ ($P=0$) is used in the cases of both the BL
and KS coordinates.
First, we show the result of the BL coordinates.
When $I=0$ ($P=0$) in the BL coordinates of $a_\ast = 0.01$,
we have $B^{\hat{\phi}}=0$.
Figure \ref{kerr3tvdk} shows that no quantity ever changes at the horizon due to
the time-freezing at the horizon. A wave is caused around
the ergosphere, and the fronts propagates outward like 
a ``tsunami" and inward to the horizon.
The constants of the stationary state in the wave are realized as the Blandford--Znajek
solution with $\Omega_{\rm F} = \omega_{\rm H}/2=a_\ast/8M=0.00125$. 
The energy flux is provided from the front of the inward wave near the horizon.
In the front of the inward wave, the energy-at-infinity becomes negative.
%
To observe this effect,
we check the above statement near the black hole, as shown in Fig. \ref{kerr3tvdkt8}.
We find the profile of $P$ at $t=4M$ has its maximum value at $r=2.2M$, and its value
at the horizon is zero. Comparing the profile of $P$ at $t=4M$ and $t=8M$
indicates that the profile spreads both outward and inward. The outward spread
is shown in Fig. \ref{kerr3tvdk}.
Note that the inward spread continue to approach to the horizon but
never actually reaches it. Then the energy flux emerges
from the ergosphere. As shown in Fig. \ref{kerr3tvdk},
the energy-at-infinity $e^\infty$ becomes negative at $t=40M$.

\begin{figure}[h!]
\includegraphics[width=16cm]{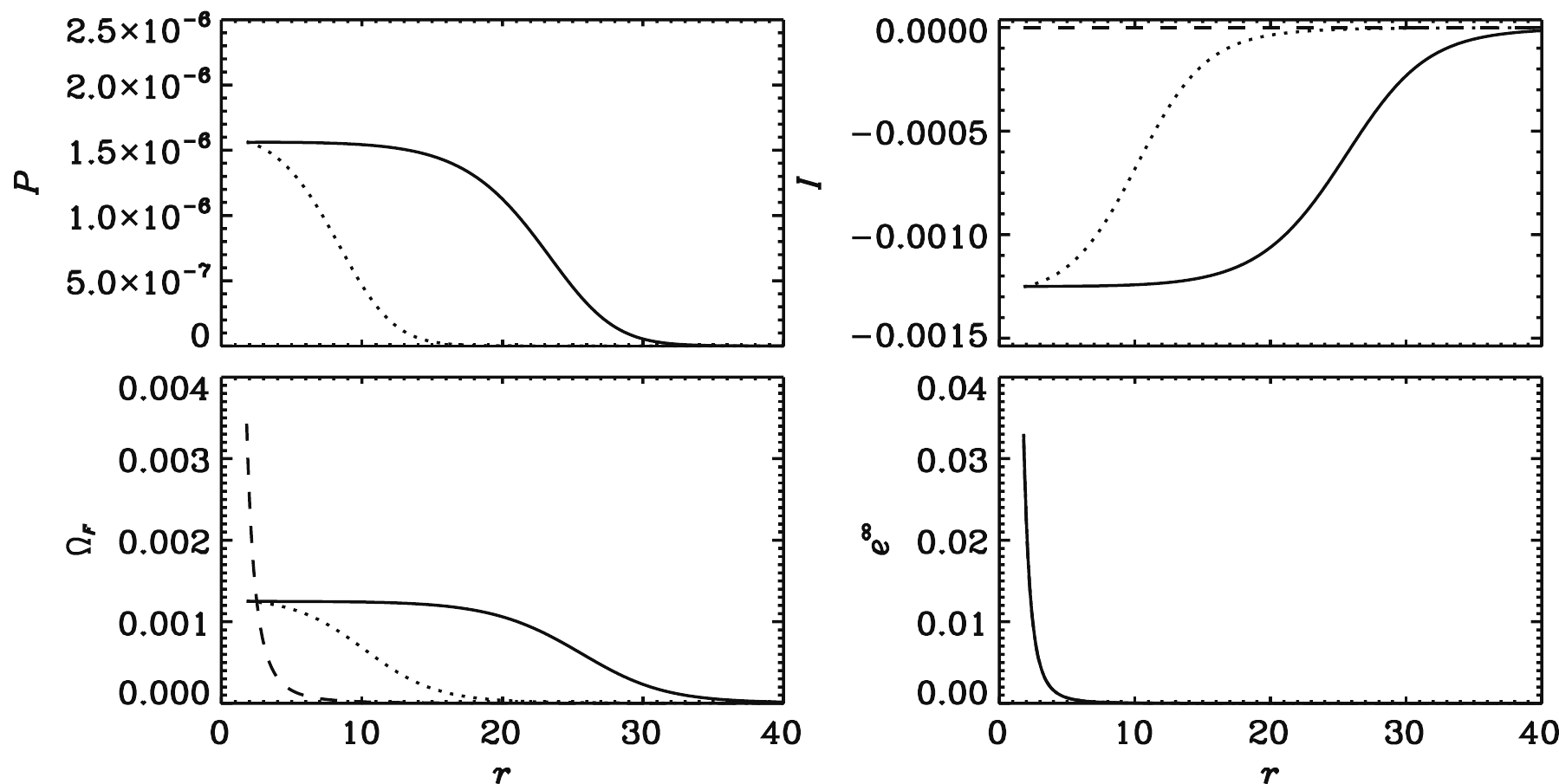}
\caption{
Simulation of FFMD with zero power and current
($P=0$, $I=0$) as initial conditions along the radial magnetic field
at the equatorial plane in terms of the KS coordinates 
($a_\ast=0.01$). 
Here the electric field in the normal frame is initially zero 
($E_{\tilde{\theta}} = 0$).
Dashed lines show quantities at $t=0$,
dotted lines show results at $t=20M$, and
solid lines show results at $t=40M$.
\label{bzd1}}
\end{figure}

In the KS coordinates, we have the initial condition $I=0$ ($P=0$) and
$E_{\tilde{\theta}} = 0$.
The results are shown in Fig. \ref{bzd1}.
The quantities at the horizon rapidly converge to the values of Blandford--Znajek
solution, and the region similar to the Blandford--Znajek solution
spreads outward. It appears that the horizon behaves like a boundary,
while ``causality" prohibits outward propagation of the information through the horizon.
This result also demonstrates that 
the outward energy flux at the
horizon increases spontaneously to reach the value
of the Blandford--Znajek solution
even in the case of initially no current and energy flux. 

\subsubsection{Test calculations of the Blandford--Znajek mechanism with finite $a_\ast$
\label{rad95stat}}




We perform 1D FFMD numerical simulations with 
the steady-state solution for finite $a_\ast$ at
the equatorial plane in the flux coordinates based on the BL
and KS coordinates. 
In the force-free steady state, we have constants
$I$, $P$, and $\Omega_{\rm F}$, and 
the azimuthal magnetic field component and colatitude electric field component 
are given by Eqs.  (\ref{bphi6iomg}) and (\ref{epsi6iomg}), 
respectively.

When the magnetic surface locates at the equatorial plane ($\theta_0 (r) = \pi/2$
and $K=0$), using Eqs. (\ref{omgfdtmd}) and (\ref{idtmd}) 
with $\theta_{\rm H} = \theta_\infty=\pi/2$, we obtain the following constants:
\begin{eqnarray}
&\displaystyle  \Omega_{\rm H} 
= \frac{a}{((b_\infty/b_{\rm H}) r_{\rm H} + 2 M)r_{\rm H}}, 
\label{omgf4stestagen} \\
&\displaystyle  I = - \frac{a b_\infty}{((b_\infty/b_{\rm H}) r_{\rm H} + 2 M)r_{\rm H}}.
\label{cur4stestagen}
\end{eqnarray}

\begin{figure}[h!]
\includegraphics[width=17cm]{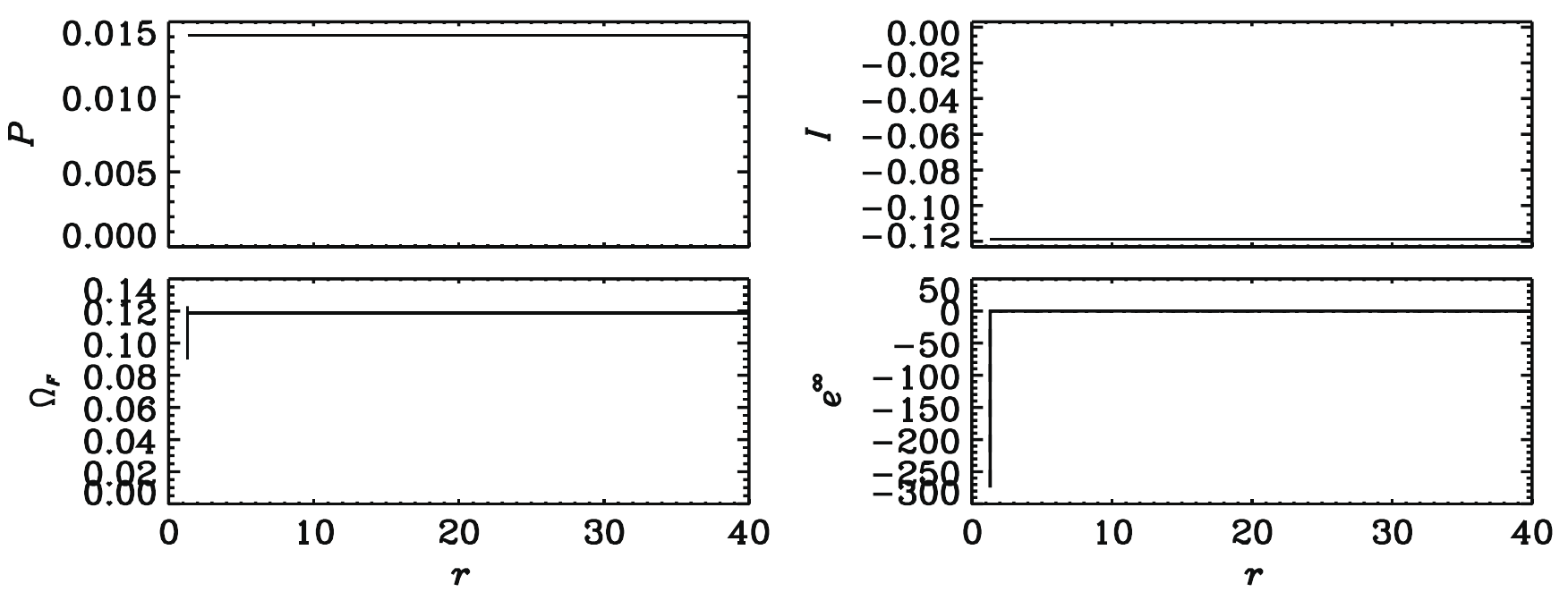}
\caption{Test by the Blandford--Znajek solution ($a_\ast=0.95$) with the BL
coordinates. Dashed lines show quantities at $t=0$,
dotted lines show results at $t=5M$, and
solid lines show results at $t=10M$.
Dashed, dotted, and solid lines overlap in this case.
\label{bzblk_a95}}
\end{figure}

\begin{figure}[h!]
\includegraphics[width=16cm]{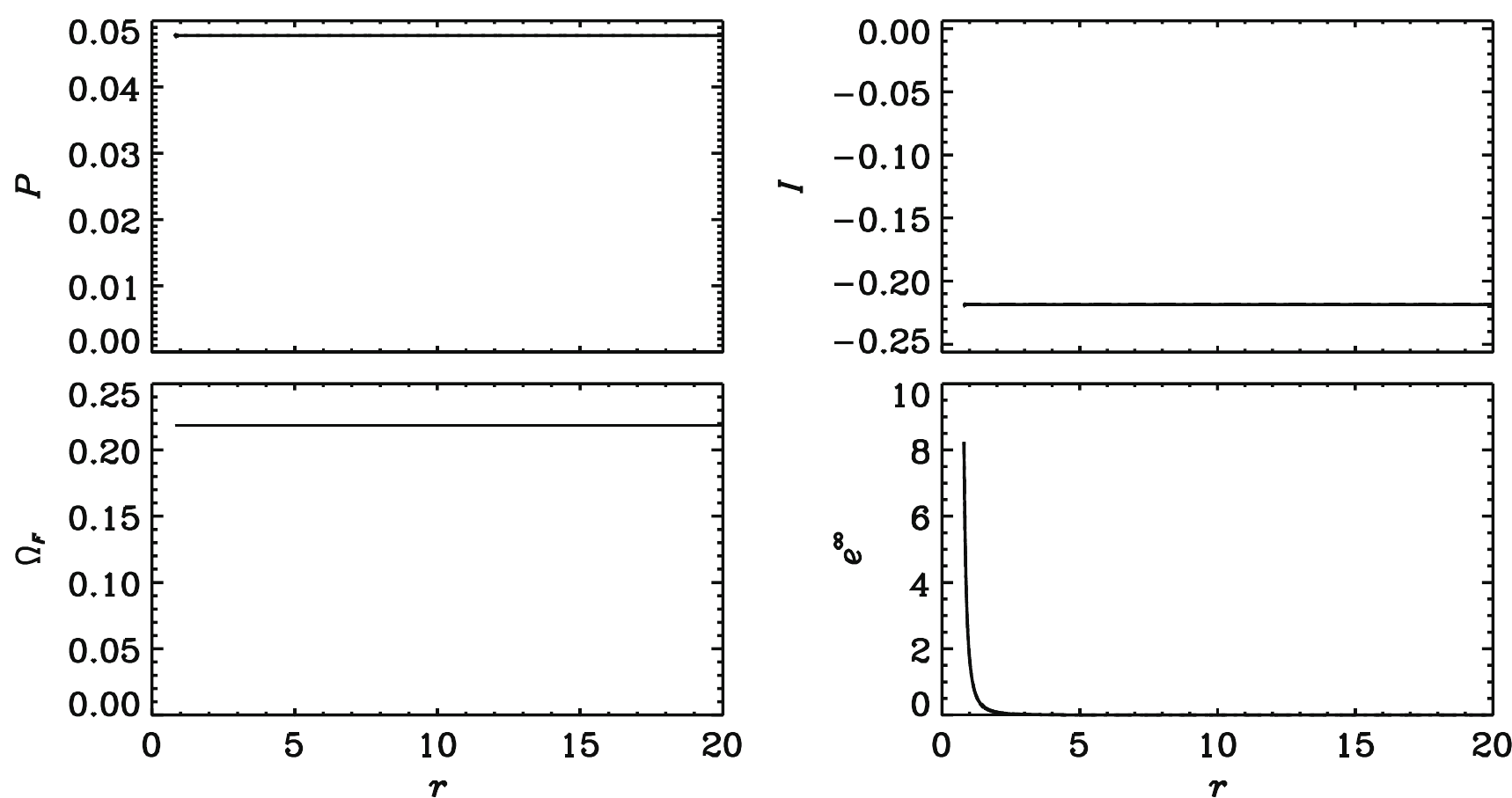}
\caption{
Simulation with steady state solution 
with Eqs. (\ref{omgf4stestagen}) and (\ref{cur4stestagen})
for $a_\ast =0.95$.
Dashed lines show quantities at $t=0$,
dotted lines show results at $t=5M$, and
solid lines show results at $t=10M$.
Dashed, dotted, and solid lines overlap in this case.
\label{finpowsol}}
\end{figure}

Figures \ref{bzblk_a95} and
\ref{finpowsol} show the simulation results with the steady-state solution 
(Eqs. (\ref{omgf4stestagen}) and (\ref{cur4stestagen}))
for the case of a radial magnetic surface ($b_\infty = b_{\rm H}$), 
where the flux coordinates are
the BL and KS coordinates, respectively, for $a_\ast = 0.95$.
Here the horizon locates at $r=1.31M$. This clearly confirms that the solution
gives the steady state of the electromagnetic field with the outward power predicted
by $P = \Omega_{\rm F}^2 = 0.048$, where we use the unit system such that
$b_{\rm H} = 1$. Hereafter, we use the unit system with $b_{\rm H} = 1$.

\subsubsection{Test calculations of pulse propagation on the Blandford--Znajek solution
with finite $a_\ast$ \label{rad95puls}}

To investigate causality, we follow the propagation of the pulse
initially added to the steady-state field around a rapidly spinning black hole
($a_\ast = 0.95$). Here we initially add the perturbation of the pulse to the
azimuthal component of the magnetic field.
The additional azimuthal component of the magnetic field of the
pulse to the stationary fields 
at $r = r_0$ with width $\delta$ and amplitude $A$ is given by
\begin{eqnarray} 
B^{\tilde{\phi}}_{\rm pul} = 
\left \{ 
\begin{array}{cl}
\displaystyle \frac{A}{2} \left ( 1 + \cos \pi \frac{r-r_0}{\delta/2} \right )
&\displaystyle  \hspace{1cm}  \left (r_0 - \frac{\delta}{2} \le r \le r_0 + \frac{\delta}{2} \right ) \\
0 &\displaystyle  \hspace{1cm}  \left (r < r_0 - \frac{\delta}{2}, r > r_0 + \frac{\delta}{2} \right )
\end{array} 
\right . .
\label{eq4pulse}
\end{eqnarray}

\begin{figure}[h!]
\includegraphics[width=16cm]{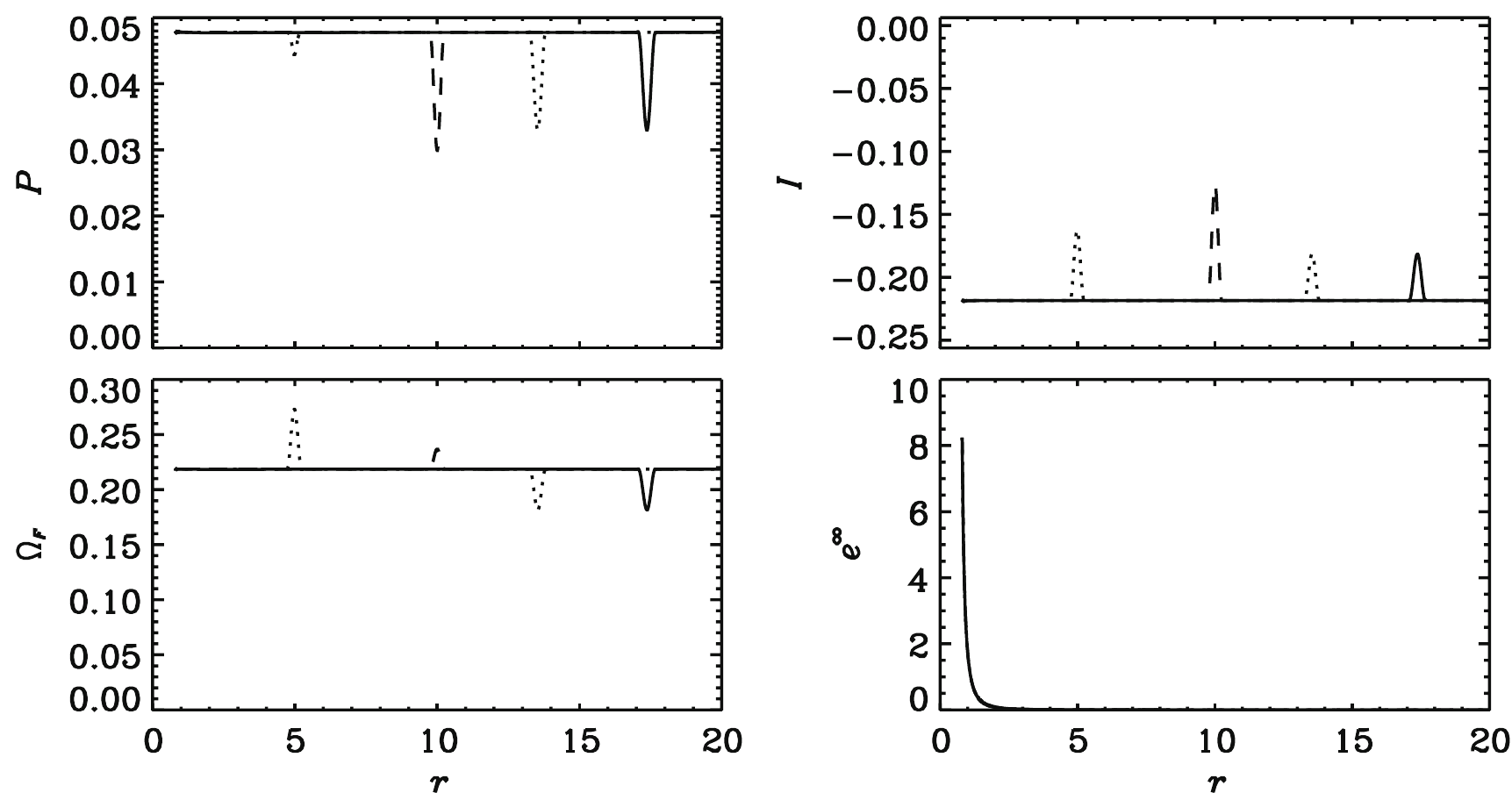}
\caption{
Simulation of the propagation of the pulse of the magnetic field
with $A=10^{-3}$ of Eq. (\ref{eq4pulse})
at $r=10$ outside the ergosphere. 
The background field is given by the steady-state solution 
with Eqs. (\ref{omgf4stestagen}) and (\ref{cur4stestagen}) 
($b_\infty/b_{\rm H} =1$ (radial magnetic surface), $a_\ast=0.95$).
Dashed lines show quantities at $t=0$,
dotted lines show results at $t=5M$, and
solid lines show results at $t=10M$.
\label{fipoads_far}}
\end{figure}

Figure \ref{fipoads_far} shows the propagation of the pulse initially at $r=10M$
on the background steady-state electromagnetic field as obtained by
Eqs.  (\ref{omgf4stestagen}) and (\ref{cur4stestagen}) 
for $a_\ast = 0.95$. Two pulses propagate outward and inward with velocity 
0.7 and 1.2, respectively. This demonstrates that 
the information can propagate both inward and
outward at near the speed of light outside of the ergosphere.

\begin{figure}[h!]
\includegraphics[width=16cm]{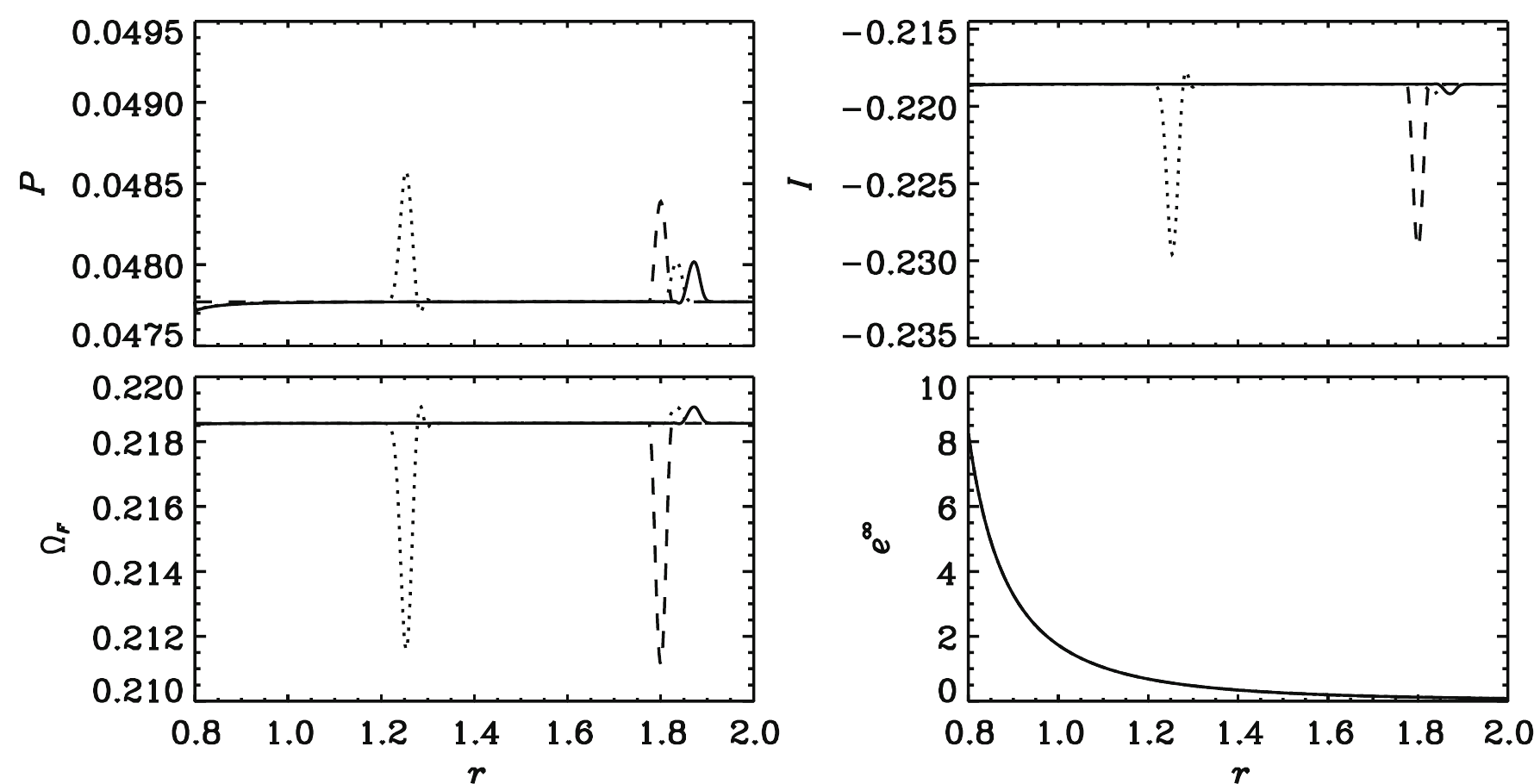}
\caption{
Simulation of the propagation of the pulse of the magnetic field
with $A=3 \times 10^{-3}$ of Eq. (\ref{eq4pulse})
at $r=1.8$ inside the ergosphere. 
The background field is given by the steady-state solution 
with Eqs. (\ref{omgf4stestagen}) and (\ref{cur4stestagen}) 
($b_\infty/b_{\rm H} =1$ (radial magnetic surface), $a_\ast=0.95$).
Dashed lines show quantities at $t=0$,
dotted lines show results at $t=1M$, and
solid lines show results at $t=2M$.
\label{fipoads_ergo}}
\end{figure}

Figure \ref{fipoads_ergo} shows the propagation of the pulse initially located at $r=1.8M$
inside the ergosphere, where the background is the same as that in Fig. \ref{finpowsol}.
As shown, two pulses also propagate outward and inward; however, the inward pulse runs fast,
and the outward pulse propagates very slowly.

\begin{figure}[h!]
\includegraphics[width=16cm]{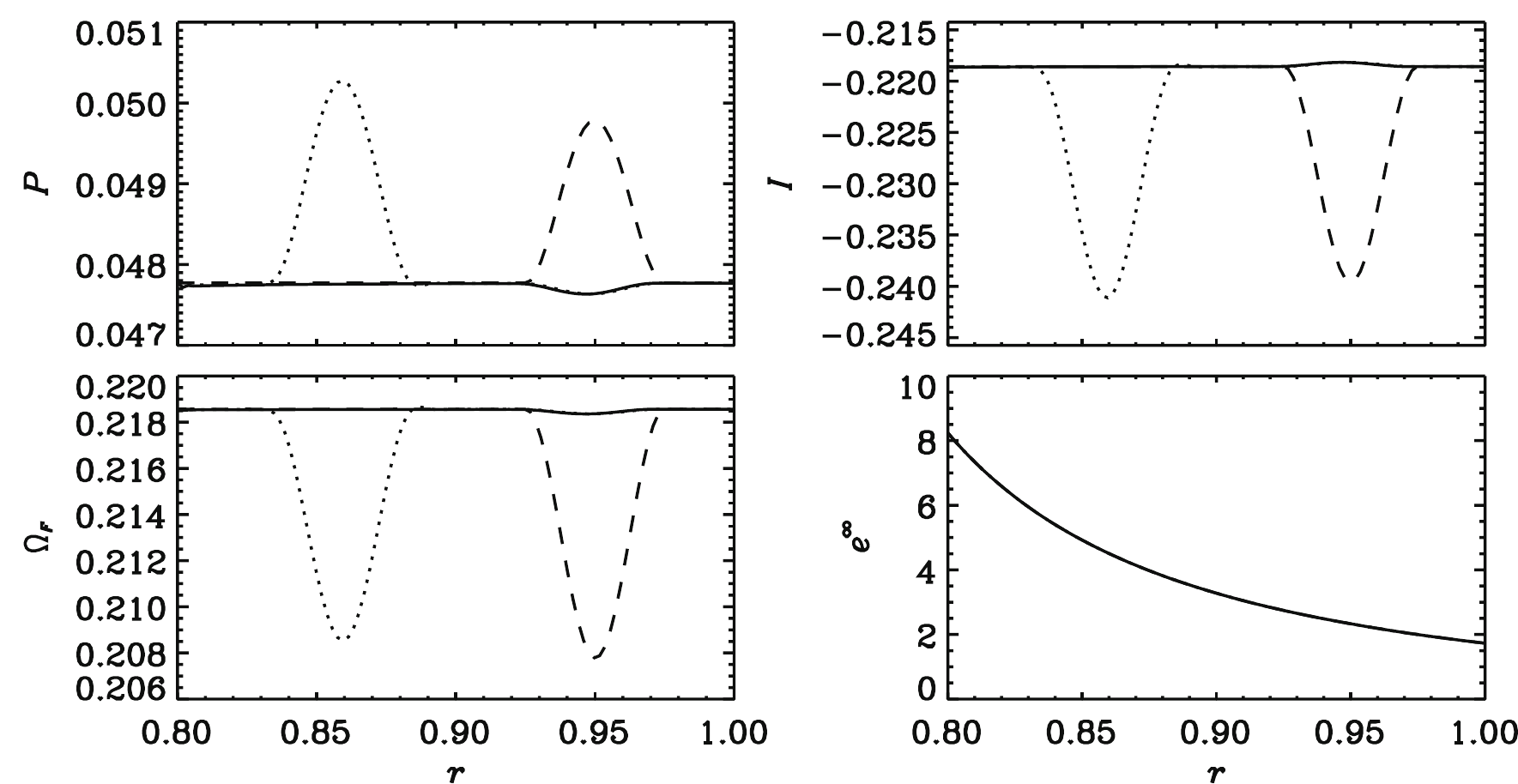}
\caption{
Simulation of the propagation of the pulse of the magnetic field
with $A=10^{-2}$ of Eq. (\ref{eq4pulse})
at $r=0.9$ inside the horizon. 
The background field is given by the steady-state solution 
with Eqs. (\ref{omgf4stestagen}) and (\ref{cur4stestagen})  
($b_\infty/b_{\rm H} =1$ (radial magnetic surface), $a_\ast=0.95$).
Dashed lines show quantities at $t=0$,
dotted lines show results at $t=0.1M$, and
solid lines show results at $t=0.2M$.
\label{fipoads_ihz}}
\end{figure}

Figure \ref{fipoads_ihz} shows the propagation of the pulse initially located at $r=0.95M$
inside the horizon, where the background is the same as that in Fig. \ref{finpowsol}.
Here we find two pulses at $t=0.1M$ (dotted line).
One pulse at $r=0.82M$ propagates inward rapidly. 
At $t=0.2M$, this pulse passes through the horizon. 
The pulse at $r = 0.945$ propagates 
inward very slowly and appears to be nearly at rest.
This confirms that 
the energy can be transported outward even inside of the horizon,
but information is never transported
outward. This result is consistent with the notion of causality around a black hole.

\subsubsection{Test calculations of the emerging Blandford--Znajek process
with finite $a_\ast$ \label{rad95tsun}}

To investigate the emergence process of the Blandford--Znajek mechanism, 
we run several simulations with the KS coordinates
and observe the time-dependent process of the force-free field inside, at, and outside
the horizon. To find the details of the process between the static limit
and the horizon, we put the spin parameter of the black hole as $a_\ast = 0.95$.
We also perform simulations with the BL coordinates
to investigate the dependence on the coordinate systems.
Here we use the initial condition with $B^{\tilde{\phi}} = 0$ and $E_{\tilde{\theta}}=0$
around the spinning black hole.

\begin{figure}[h!]
\includegraphics[width=16cm]{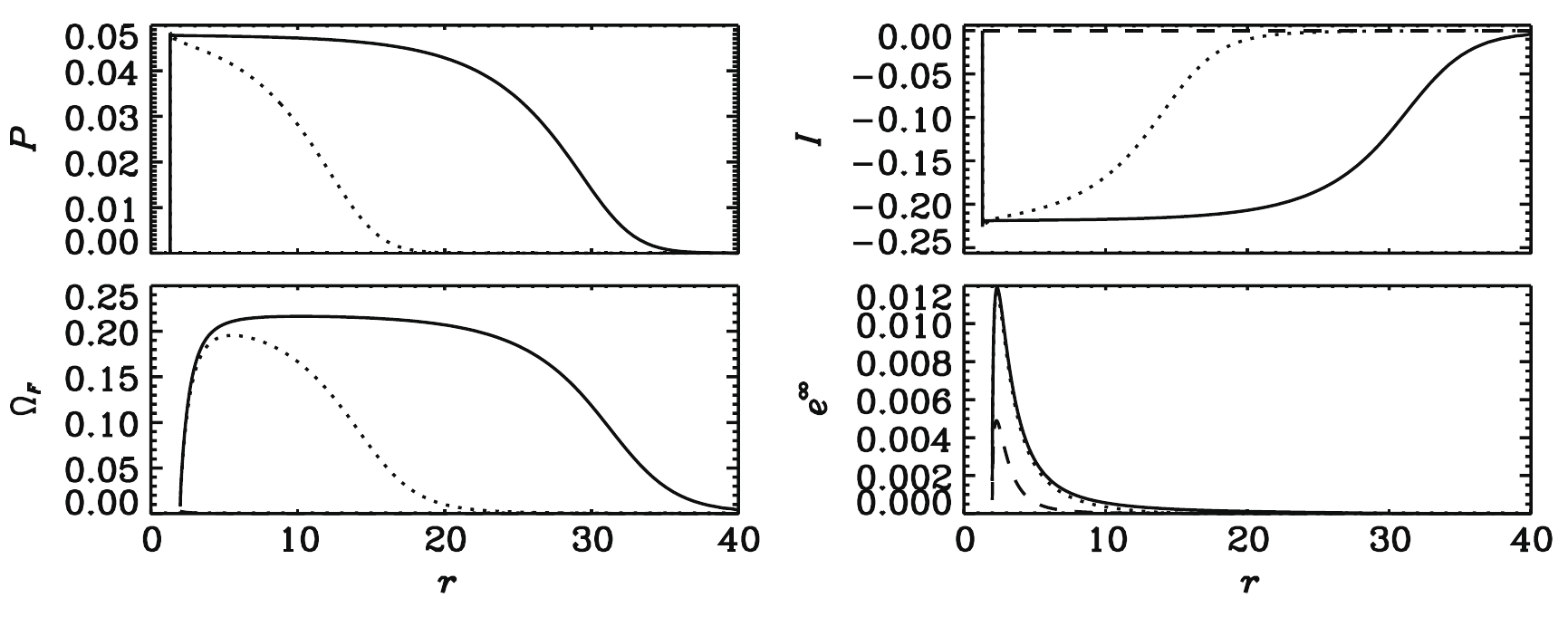}
\caption{
Simulation of FFMD with zero power and current
(initial condition: $P=0$, $I=0$) 
as initial conditions in terms of the BL coordinates ($a_\ast=0.95$). 
Dashed lines show quantities at $t=0$,
dotted lines show results at $t=20M$, and
solid lines show results at $t=40M$.
\label{kerr3tvdk_a95t40}}
\end{figure}

\begin{figure}[h!]
\includegraphics[width=16cm]{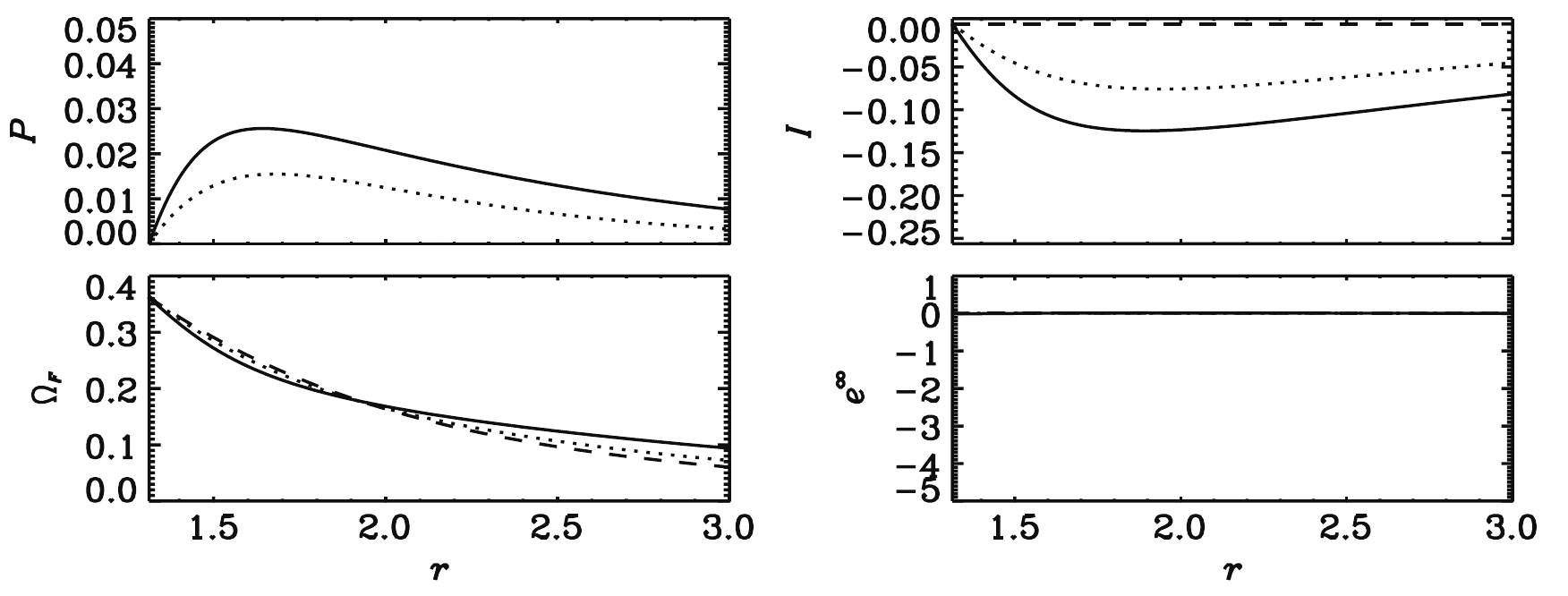}
\caption{
Enlarged plot near the horizon at an early stage of the simulation 
(initial conditions: $P=0$, $I=0$) of Fig. \ref{kerr3tvdk}
in terms of the BL coordinates ($a_\ast=0.95$). 
Dashed lines show quantities at $t=0$,
dotted lines show results at $t=4M$, and
solid lines show results at $t=8M$.
\label{kerr3tvdk_a95nea}}
\end{figure}

Figure \ref{kerr3tvdk_a95t40} shows the time development of the power $P$,
current $I$, and angular velocity of the magnetic field lines $\Omega_{\rm F}$
on the BL coordinates.
The power and angular velocity of the magnetic field lines are
zero at the initial condition. The results obtained at $t=20M$ (dotted line) and $t=40M$
(solid line) show that finite regions of $P$ and $\Omega_{\rm F}$ spread outward
at the speed of light. Note that $P$ and $\Omega_{\rm F}$ converge to
the values expected by the steady state, i.e., $P = \Omega_{\rm F}^2 = 0.048$ 
with Eqs. (\ref{omgf4stestagen}) and (\ref{cur4stestagen}). 
Then it seems that the steady-state outward 
energy flux is supplied at the vicinity of the black hole.
Figure \ref{kerr3tvdk_a95nea} shows an enlargement of Fig. \ref{kerr3tvdk_a95t40}
near the horizon. Here we observe inward propagation of the tsunami of $P$ and 
$\Omega_{\rm F}$ in the BL coordinates, which is caused by
a time lapse of the coordinates near the horizon.

\begin{figure}[h!]
\includegraphics[width=16cm]{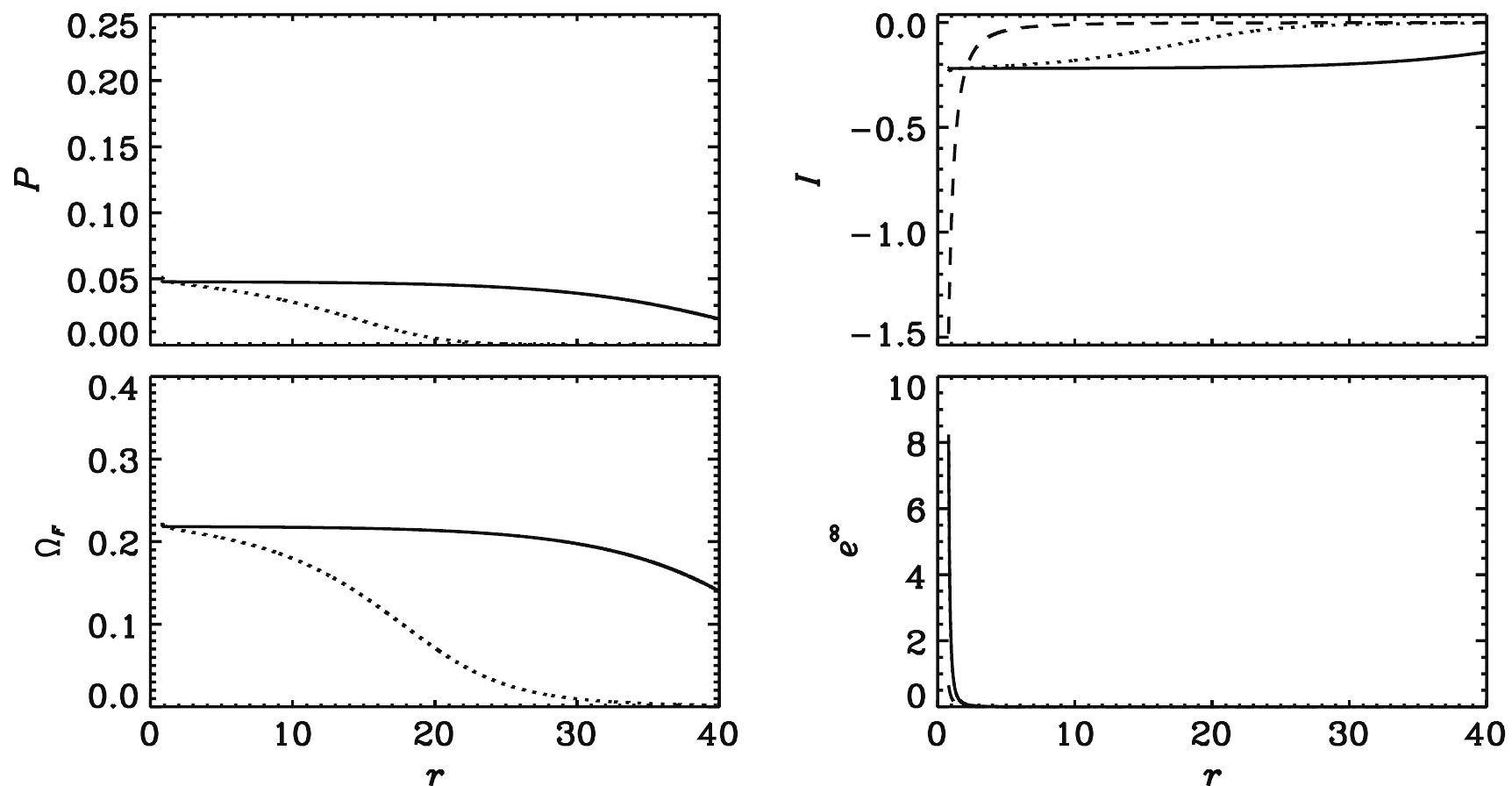}
\caption{
Simulation of the case with initial conditions of
$B^{\tilde{\phi}}=0$, $E_{\tilde{\theta}}=0$ 
along the radial magnetic surface at the equatorial plane around a rapidly spinning 
black hole ($a_\ast=0.95$).
Dashed lines show quantities at $t=0$,
dotted lines show results at $t=30M$, and
solid lines show results at $t=60M$.
\label{allzero60}}
\end{figure}

Figure \ref{allzero60} shows the time development of $P$, $I$, and $\Omega_{\rm F}$
on the KS coordinates.
The power and angular velocity of the magnetic field lines are
zero in the initial condition. The results at $t=30M$ (dotted line) and $t=60M$
(solid line) show that finite regions of $P$ and $\Omega_{\rm F}$ spread outward
at the speed of light. Note that $P$ and $\Omega_{\rm F}$ converge to
the values expected by the steady state, i.e., $P = \Omega_{\rm F}^2 = 0.048$ 
with Eqs.  (\ref{omgf4stestagen}) and (\ref{cur4stestagen}). 
Then it seems that the steady-state outward 
energy flux comes form the vicinity of the black hole.

\subsubsection{Test calculations of the emergence of the Blandford--Znajek mechanism
along the incurvature-flared magnetic surface at the equatorial plane
\label{par95}}

Here we show a numerical calculation with an incurvature-flared magnetic surface 
given by Eq. (\ref{eq4q}) with $m = 0.25$
imaged in Fig. \ref{variousmagsurf} (c).

\begin{figure}[h!]
\includegraphics[width=16cm]{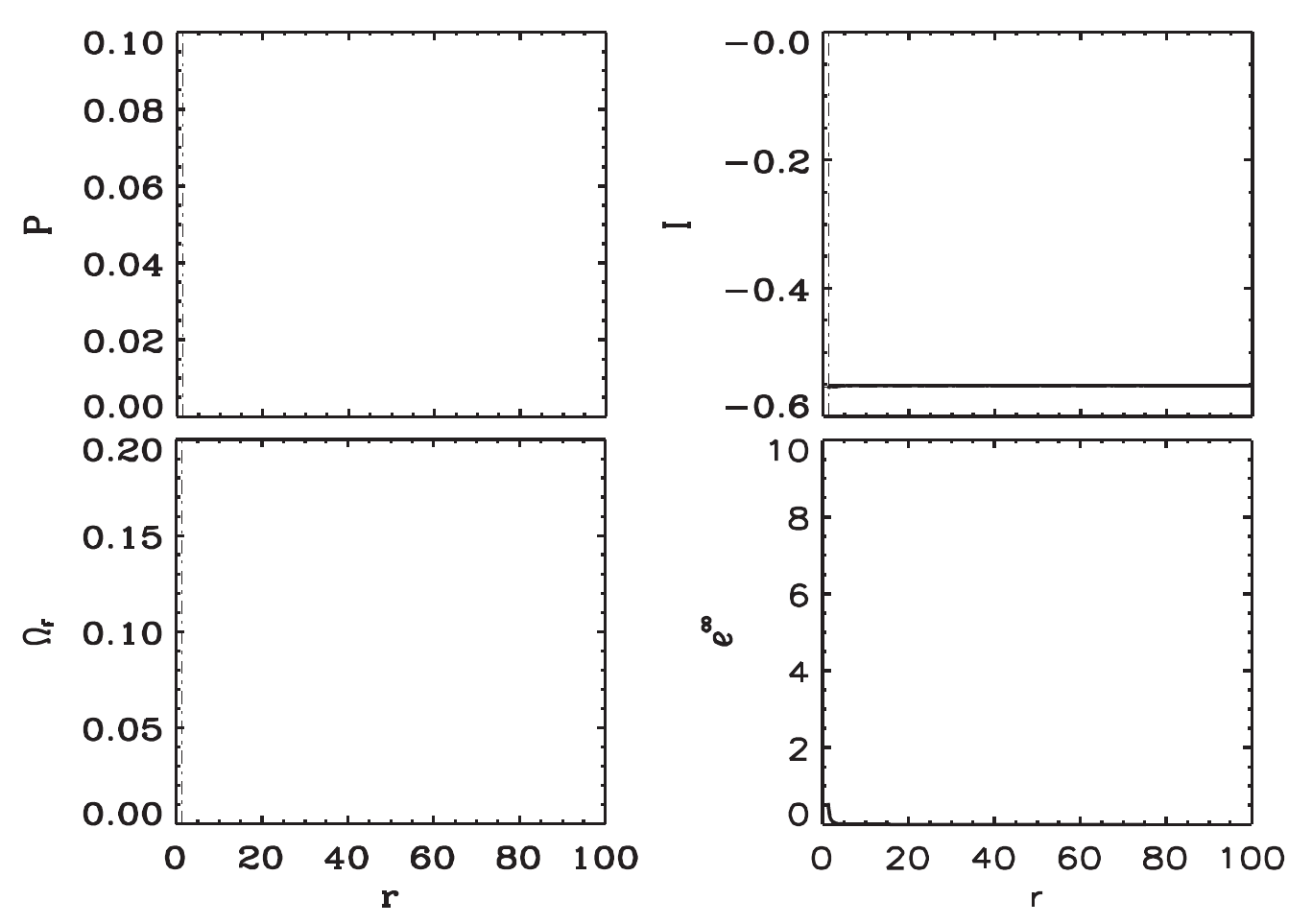}
\caption{
Simulation of the case with initial conditions of 
a steady-state solution of the force-free electromagnetic field
along the incurvature-flared magnetic surface at the equatorial plane ($m=0.25$)
around a rapidly spinning black hole ($a_\ast=0.95$).
Dashed lines show quantities at $t=0$,
dotted lines show results at $t=5M$, and
solid lines show results at $t=10M$.
Dashed, dotted, and solid lines overlap in this case.
\label{parabostat}}
\end{figure}

Figure \ref{parabostat} shows the steady-state solution of $P$, $I$, and $\Omega_{\rm F}$
along an incurvature-flared magnetic surface
at the equatorial plane around a spinning black hole with $a_\ast = 0.95$.

\begin{figure}[h!]
\includegraphics[width=18cm]{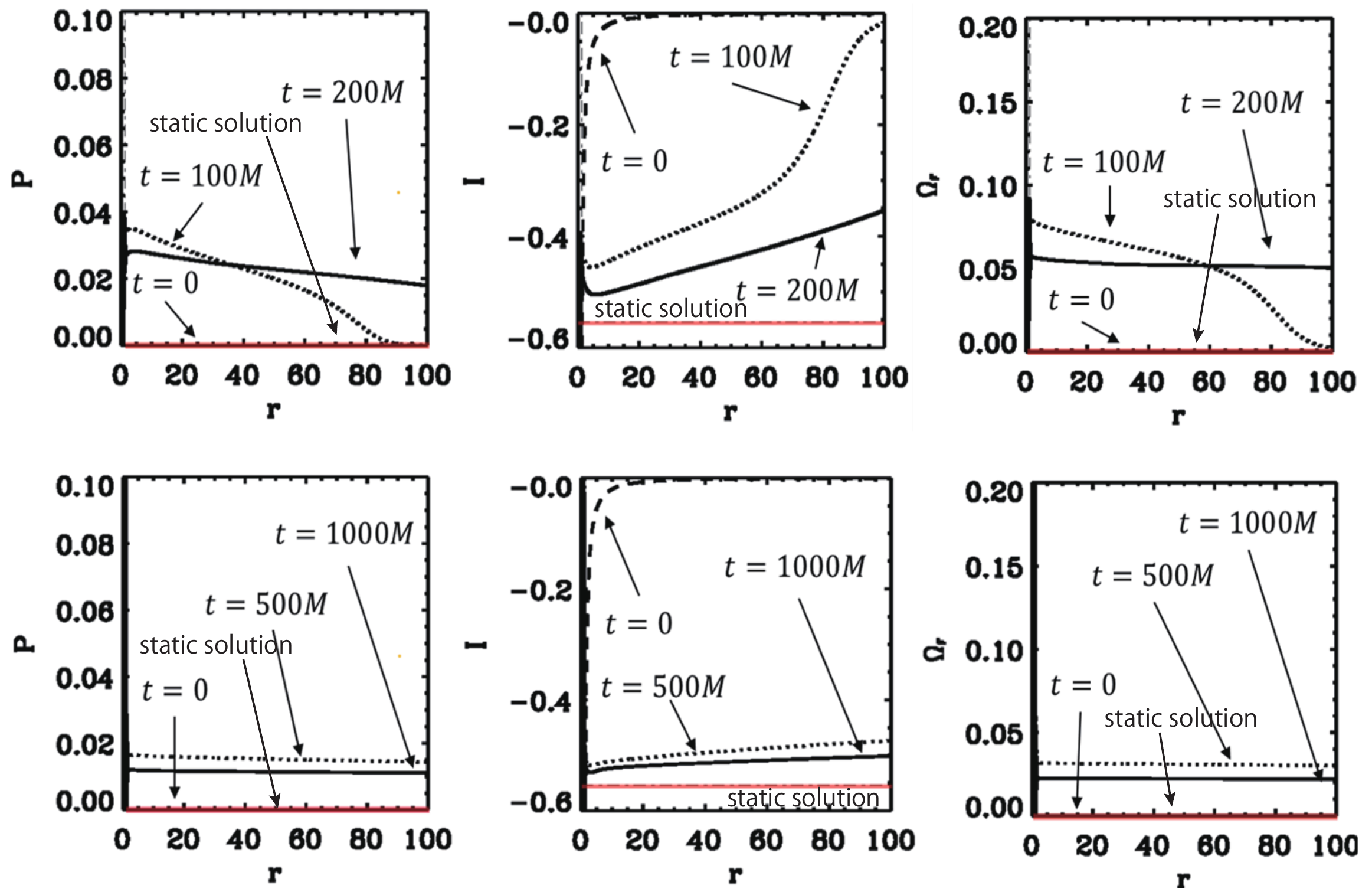}
\caption{
Simulation of the case with initial conditions of 
$B^{\tilde{\phi}}=0$, $E_{\tilde{\theta}}=0$ 
along the incurvature-flared magnetic surface at the equatorial plane
($m=0.25$) around a rapidly spinning black hole ($a_\ast=0.95$).
Dashed lines show quantities at $t=0$.
Horizontal solid red lines show static solutions.
The top panels show results early stages 
(dotted lines: $t=100M$; solid lines: $t=200M$).
The bottom panels show results at late stages
(dotted lines: $t=1000M$; solid lines: $t=2000M$).
\label{parabo}}
\end{figure}

Figure \ref{parabo} shows the time development of $P$, $I$, and $\Omega_{\rm F}$
of the force-free electromagnetic field along an incurvature-flared magnetic surface
at the equatorial plane around a spinning black hole with $a_\ast = 0.95$.
Here the horizontal solid red lines show the steady-state solution.
In the incurvature-flared magnetic surface case, the electromagnetic field
tends to converge very gradually to the steady state 
(details are provided by \citealt{imamura19}).

\subsubsection{Test calculations of the emergence of the Blandford--Znajek mechanism
along the excurvature-flared magnetic surface at the equatorial plane
\label{qua95}}

\begin{figure}[h!]
\includegraphics[width=18cm]{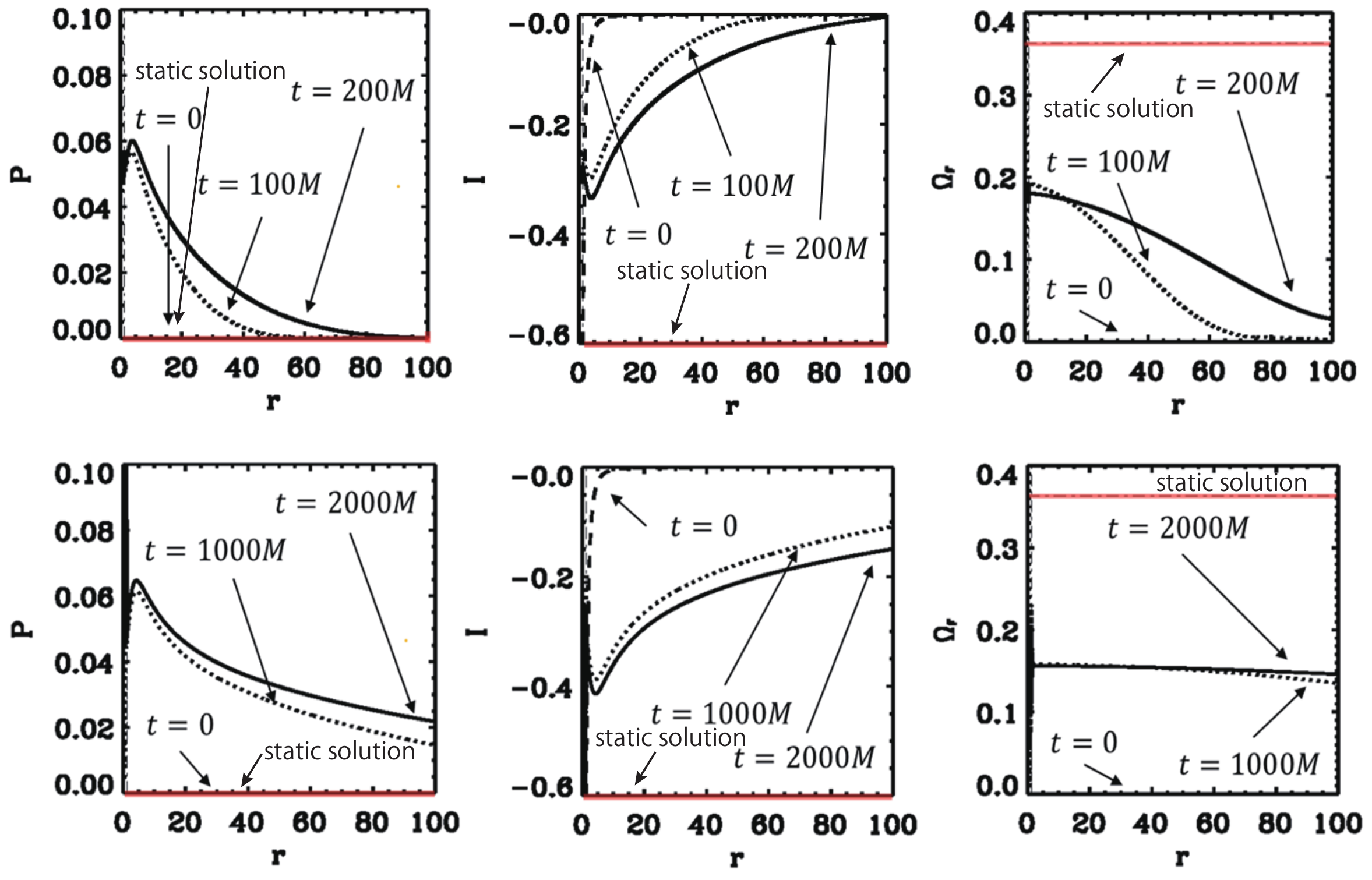}
\caption{
Simulation of the case with initial conditions of 
$B^{\tilde{\phi}}=0$, $E_{\tilde{\theta}}=0$ 
along the excurvature-flared magnetic surface at the equatorial plane
($m=-0.25$) around a rapidly spinning black hole ($a_\ast=0.95$).
Dashed lines show quantities at $t=0$.
Horizontal solid red lines show static solutions.
The top panels show results at early stages 
(dotted lines: $t=100M$; solid lines: $t=200M$).
The bottom panels show results at late stages
(dotted lines: $t=1000M$; solid lines: $t=2000M$).
\label{quadrapole}}
\end{figure}

Here we show a numerical calculation with an excurvature-flared magnetic surface
given by Eq. (\ref{eq4q}) with $m=-0.25$ imaged in Fig. \ref{variousmagsurf} (b).
Figure \ref{quadrapole} shows the time development of $P$, $I$, and $\Omega_{\rm F}$
of the force-free electromagnetic field along an excurvature-flared magnetic surface
at the equatorial plane around a spinning black hole with $a_\ast = 0.95$.
The horizontal solid red lines show the steady state solution.
In the excurvature-flared magnetic surface case, the electromagnetic field
does not appear to converge to the steady state
(details are given by \citealt{imamura19}).

\section{Summary \label{secdiscussion}}

We have reviewed the basic method of 1D FFMD and showed the analytic solution of the 
steady-state force-free field for an arbitrary magnetic surface with
Eqs. 
(\ref{bphi6iomg}), (\ref{epsi6iomg}), (\ref{ieqqsinqomgf}), and (\ref{omgfdtmd}).
To confirm the validity of the analytic solution, we derived two
analytic solutions given by \citet{blandford77} using 
Eqs. (\ref{omgfdtmd}) and (\ref{idtmd}).
We have demonstrated test numerical calculations of 1D FFMD for three types of
magnetic surfaces along the equatorial plane (in Table 1).
The numerical simulations of 1D FFMD are used to investigate the
mechanism of energy extraction from a spinning black hole
via the magnetic field and transport of the 
extracted energy toward infinity \citep{koide18,imamura19}.
First, in the case of a slowly spinning black hole ($a_\ast \ll 1$),
we have confirmed the analytic solution for the
radial magnetic surface of a magnetic monopole derived by
\citet{blandford77} with both the BL and KS
coordinates. We have also performed simulations of nonstationary fields
on the magnetic surface along the equatorial plane around
rapidly spinning black holes.
The electromagnetic field approaches those of
the Blandford--Znajek solution spontaneously, except for
the following excurvature-flared magnetic surface cases.
With the radial magnetic surface in the KS coordinates, 
under any initial conditions,
the quantities at the horizon converge rapidly to the value
given by the Blandford--Znajek solution, and the region
of the Blandford--Znajek solution spreads toward infinity.
Conversely, with the radial magnetic surface in 
the BL coordinates, except for the case of the
initial condition of the Znajek condition at the horizon, no
energy radiates at the horizon due to time-freezing at the horizon. 
The tsunami-like wave spreads outward
to infinity and inward to the horizon. In the tsunami-like
wave region, the Poynting flux is directed outward. At the front of the
wave near the horizon, energy-at-infinity becomes significantly negative 
to provide energy to the tsunami-like outward wave. In other words, the energy source
of the outward Poynting flux is the negative energy region 
at the front of the inward wave near the horizon.
This clearly demonstrates the difference between the results 
obtained in the BL
and KS coordinates. The numerical results of the BL
coordinates become the same as those of the KS coordinates
distant from but not near the horizon. 
This difference is reasonable because the time is
different between the BL and KS coordinates 
at the horizon and infinitely distant from the horizon (Eq. (\ref{rel4bl2ks})). 
In other words, the initial condition of the BL 
coordinate frames at the horizon corresponds to the condition
at infinite past in the KS coordinates.
In other cases with incurvature- and excurvature-flared magnetic surfaces,
the regions of finite power density ($P > 0$), where $P$ is initially zero,
spread gradually outward. In the incurvature-flared magnetic surface case,
the electromagnetic field state ($\Omega_{\rm F}$, $I$, and $P$)
converges to the expected solution of the steady-state solution 
after a very long period ($t=1000M$). In the 
excurvature-flared magnetic surface case, the electromagnetic field state
($\Omega_{\rm F}$, $I$, and $P$) appears not to converge
to the steady state. Here the steady state is never
achieved. Detailed analyses of incurvature- and excurvature-flared
magnetic surfaces have been performed previously by \citet{imamura19}.

The results shown in Figs. \ref{finpowsol} -- \ref{fipoads_ihz} provide important
insight into the causality relative to the Blandford--Znajek mechanism.
These results suggest that there are two kinds of energy flux with the information.
One kind of energy flux can bring information in the direction of the energy flux,
similar to a pulse outside the horizon.
The other kind of energy flux never brings information 
in the direction of the outward energy flux, such as a pulse inside the horizon.
This appears to explain causality in the Blandford--Znajek mechanism.


In conclusion, we have reviewed the basic method of 1D FFMD, 
given the analytic solutions of the 
steady-state force-free field for an arbitrary magnetic surface,
and demonstrated test calculations of 1D FFMD with three types of 
magnetic surfaces at the equatorial plane.
Using the analytic solutions of 1D FFMD, we may be able to analytically 
solve the Grad--Shafranov equation
relative to $\Psi(r, \theta)$ around a spinning black hole.
This may become a remarkable challenge.
We expect that 1D FFMD will be employed frequently as a standard tool in various fields.


\begin{acknowledgments}
S.K. thanks Mika Koide for her useful comments on this paper.
We also thank Masaaki Takahashi and Fumio Takahara 
for their plentiful discussion and suggestions on this study.
\end{acknowledgments}

\appendix

\section{3+1 formalism of energy and momentum conservation laws in 1D FFMD \label{appenda}}

We show the derivations of the 3+1 formalism of 
conservation laws of energy and momentum in
FFMD (Eqs. (\ref{3+1enem}) and (\ref{3+1moem})).
The world line of the normal observer frame is perpendicular to the
hypersurface of constant time in spacetime. The four-velocity of the normal observer frame
is $N^\mu = (1/\alpha, - \beta^i/\alpha)$, $N_\mu = (-\alpha, 0)$.
The projection operator to a space-like hypersurface is written by ${\cal P}_{\mu\nu}
= g_{\mu\nu} + N_\mu N_\nu$. The temporal component of an arbitrary vector $F^\mu$ 
observed by the normal observer frame is 
\begin{equation}
\tilde{F}^{\dagger} = F^\mu N_\mu.
\end{equation}
The projection of the vector $F^\mu$ to the space-like hypersurface
\begin{equation}
\tilde{F}^{\mu} = {\cal P}^{\mu\nu} F_\nu ,
\end{equation}
and $\tilde{F}^{\dagger}$ produces $F^\mu$ as
\begin{equation}
F^\mu = \tilde{F}^\mu + \tilde{F}^{\dagger} N^\mu.
\end{equation}
Components of the vector measured by the normal observer frame
are written by $F^{\tilde{\mu}} = (\tilde{F}^{\dagger}, \tilde{F}^i)$.
With respect to an arbitrary tensor $T^{\mu\nu}$, using
\begin{eqnarray}
\tilde{T}^{\dagger \dagger} = T^{\rho\sigma} N_\rho N_\sigma,
\tilde{T}^{\dagger \nu} = T_{\rho\sigma} N^\rho {\cal P}^{\sigma \nu},
\tilde{T}^{\mu \dagger} = T_{\rho\sigma} {\cal P}^{\rho \mu} N^\sigma , 
\tilde{T}^{\mu \nu} =  T_{\rho\sigma} {\cal P}^{\rho \mu} {\cal P}^{\sigma \nu}
\end{eqnarray}
we have the expansion form
\begin{equation}
T^{\mu\nu} = \tilde{T}^{\dagger \dagger} N^\mu N^\nu + \tilde{T}^{\dagger \nu} N^\mu
+ \tilde{T}^{\mu \dagger} N^\nu + \tilde{T}^{\mu \nu}.
\end{equation}
When $T^{\mu \nu}$ is a symmetric tensor, we have 
$\tilde{T}^{\dagger \mu} = \tilde{T}^{\mu \dagger}$.
When $T^{\mu \nu}$ is an antisymmetric tensor, we have $\tilde{T}^{\dagger \dagger} =0$ and
$\tilde{T}^{\dagger \mu} = -\tilde{T}^{\mu \dagger}$.
Components of the tensor measured by the normal observer frame
are written by $T^{\tilde{t} \tilde{t}} = \tilde{T}^{\dagger \dagger}$,
$T^{\tilde{t} \tilde{j}} = \tilde{T}^{{\dagger} j}$,
$T^{\tilde{i} \tilde{t}} = \tilde{T}^{i {\dagger}}$,
$T^{\tilde{i} \tilde{j}} = \tilde{T}^{ij}$.

Applying the above relations to the electromagnetic field tensor $F^{\mu\nu}$,
its dual tensor $\,^\ast F^{\mu\nu}$, and current density $J^\mu$, we have 
\begin{eqnarray}
& \displaystyle  F^{0i} = \frac{1}{\alpha} F^{\tilde{0} \tilde{i}}
= \frac{1}{\alpha} E^{\tilde{i}} , \label{rel2e4tilde} \\
& F^{ij} = {F}^{\tilde{i} \tilde{j}} + N^i \tilde{F}^{0j}
- N^j \tilde{F}^{0i}
= \epsilon^{{i}{j}{k}}
(B_{\tilde{k}} + \epsilon_{{k}{m}{n}}
N^m E^{\tilde{n}}), \\
& \displaystyle  \,^\ast F^{0i}  = \frac{1}{\alpha} \,^\ast F^{\tilde{0} \tilde{i}}
= \frac{1}{\alpha} B^{\tilde{i}} , \\
& \,^\ast F^{ij}  = - \epsilon^{ijk} (F_{\tilde{k} \tilde{0}} 
- \epsilon_{kpq} N^p \,^\ast F^{\tilde{0} \tilde{q}})
= - \epsilon^{ijk} (E_{\tilde{k}} - \epsilon_{kpq} N^p B^{\tilde{q}}), \\
& \displaystyle  J^0 = \frac{1}{\alpha} J^{\tilde{0}} = \frac{1}{\alpha} \tilde{\rho}_{\rm e}, \\
& J^i  = J^{\tilde{i}} + N^i J^{\tilde{0}} = J^{\tilde{i}}
+ N^i \tilde{\rho}_{\rm e} .
\label{rel2j4tilde}
\end{eqnarray}
Here we use $\epsilon^{\mu\nu\kappa} \equiv - N_\lambda \epsilon^{\lambda \mu \nu \kappa}
= \eta^{0\mu\nu\kappa}/\sqrt{\gamma}$.



With respect to a symmetric energy-momentum tensor $T^{\mu\nu}$,
the 3+1 formalism of the energy-momentum Eqs. (\ref{eqenmo}) and 
(\ref{eqenm0}), $\nabla_\mu T^{\mu\nu} = F^\nu$ 
is much complicated compared with that of an anti-symmetric
tensor of electromagnetic field $F^{\mu\nu}$. 
When we use $\tilde{u} = \tilde{T}^{\dagger \dagger}$, $S^{\tilde{\mu}} = \tilde{T}^{\mu \dagger}
= \tilde{T}^{{\dagger} \mu}$, $T^{\tilde{\mu} \tilde{\nu}} = \tilde{T}^{\mu\nu}$,
we obtain
\begin{eqnarray}
& \displaystyle  \alpha \tilde{F}^{\dagger} = \frac{\partial \tilde{u}}{\partial t}
+ \frac{1}{\sqrt{\gamma}} \partial_i 
\left [ \sqrt{\gamma} \alpha (S^{\tilde{i}} - N^i \tilde{u}) \right ]
+ (\partial_i \alpha) S^{\tilde{i}} - \alpha K_{ij} T^{\tilde{i} \tilde{j}} , \\
& \displaystyle  \alpha \tilde{F}_{\tilde{i}} = \frac{\partial S_{\tilde{i}}}{\partial t}
+ \frac{1}{\sqrt{\gamma}} \partial_j 
\left [ \sqrt{\gamma} \alpha (T^{\tilde{j}}_{\tilde{i}} - N^j S_{\tilde{i}}) \right ]
+ \tilde{u} \partial_i \alpha - (\partial \alpha N^j) S_{\tilde{j}}
- \frac{1}{2} \alpha (\partial_i \gamma_{jk}) T^{\tilde{j} \tilde{k}},
\end{eqnarray}
where $K_{ij} \equiv - {\cal P}^\mu_i {\cal P}^\nu_j N_{\mu ; \nu}
= - (1/2 \alpha) (\gamma_{kj} (\alpha N^k)_{,i}
+ \gamma_{ik} (\alpha N^k)_{,j} + \alpha N^k \gamma_{ij,k})$ 
is the external curvature of the spacetime.
In the case of the force-free field, setting $F^\mu = 0$, we have
\begin{eqnarray}
& \displaystyle  \frac{\partial \tilde{u}}{\partial t} =
- \frac{1}{\sqrt{\gamma}} \partial_i 
\left [ \sqrt{\gamma} \alpha (S^{\tilde{i}} - N^i \tilde{u}) \right ]
- (\partial_i \alpha) S^{\tilde{i}} + \alpha K_{ij} T^{\tilde{i} \tilde{j}} , \\
& \displaystyle  \frac{\partial S_{\tilde{i}}}{\partial t} =
- \frac{1}{\sqrt{\gamma}} \partial_j 
\left [ \sqrt{\gamma} \alpha (T^{\tilde{j}_{\tilde{i}}} - N^j S_{\tilde{i}}) \right ]
- \tilde{u} \partial_i \alpha + (\partial \alpha N^j) S_{\tilde{j}}
+ \frac{1}{2} \alpha (\partial_i \gamma_{jk}) T^{\tilde{j} \tilde{k}}.
\end{eqnarray}
Here we used $N_{i;\nu} N^\nu = \partial_i (\ln \alpha)$ and
$N_{0;\nu} N^\nu = - \alpha N^i \partial_i (\ln \alpha)$,
where the semicoron indicates the covariant derivative.


\section{Non--1D equations of FFMD \label{non1deq}}

Here we list non--1D equations of FFMD. Non--1D equations include the $\Psi$-derivative
of the electromagnetic field variables,
and information on the electromagnetic field of the neighbor magnetic surface
is required to use the non--1D equations.

The non--1D equation from Eq. (\ref{gauss2ele}) is
\begin{equation}
\tilde{\rho}_{\rm e} = \frac{1}{\sqrt{\gamma}}
\left [
\frac{\partial}{\partial r} ( \sqrt{\gamma} E^{\tilde{r}} ) 
+ \frac{\partial}{\partial \Psi} ( \sqrt{\gamma} E^{\tilde{\Psi}} ) \right ] .
\end{equation}
The non--1D equations from Eq. (\ref{ampere}) are
\begin{eqnarray}
& \displaystyle  \frac{\partial}{\partial t} E^{\tilde{r}} + \alpha (J^{\tilde{r}} +
\tilde{\rho}_{\rm e} N^r )
= \frac{1}{\sqrt{\gamma}} \frac{\partial}{\partial \Psi}
[\alpha (B_{\tilde{\phi}} + \sqrt{\gamma} (N^r E^{\tilde{\Psi}} - N^\Psi E^{\tilde{r}})],\\
& \displaystyle  \frac{\partial}{\partial t} E^{\tilde{\phi}} + \alpha (J^{\tilde{\phi}} +
\tilde{\rho}_{\rm e} N^\phi )
= \frac{1}{\sqrt{\gamma}} \frac{\partial}{\partial r}
[\alpha (B_{\tilde{\Psi}} + \sqrt{\gamma} (N^\phi E^{\tilde{r}} - N^r E^{\tilde{\phi}})]
\nonumber \\
& \displaystyle  - \frac{1}{\sqrt{\gamma}} \frac{\partial}{\partial \Psi}
[\alpha (B_{\tilde{r}} + \sqrt{\gamma} 
(N^\Psi E^{\tilde{\phi}} - N^\phi E^{\tilde{\Psi}})].
\end{eqnarray}
The non--1D equation of Eq. (\ref{3+1enem}) is
\begin{eqnarray}
& \displaystyle  \frac{\partial u}{\partial t}
= - \frac{1}{\sqrt{\gamma}} \frac{\partial}{\partial r}
[\alpha \sqrt{\gamma} (S^{\tilde{r}} + N^r \tilde{u})]
- \frac{1}{\sqrt{\gamma}} \frac{\partial}{\partial \Psi}
[\alpha \sqrt{\gamma} (S^{\tilde{\Psi}} + N^{\Psi} \tilde{u})]
\nonumber \\
& \displaystyle  - \frac{\partial \alpha}{\partial r} S^{\tilde{r}}
- \frac{\partial \alpha}{\partial \Psi} S^{\tilde{\Psi}}
\alpha K_{ij} T^{\tilde{i} \tilde{j}} .
\end{eqnarray}
Last, the non--1D equation from Eq. (\ref{3+1moem}) is
\begin{eqnarray}
& \displaystyle  \frac{\partial}{\partial t} S_{\tilde{\Psi}}
= - \frac{1}{\sqrt{\gamma}} \frac{\partial}{\partial r}
[\alpha \sqrt{\gamma} (T^{\tilde{r}}_{\tilde{\Psi}} + N^r S_{\tilde{\Psi}})]
- \frac{1}{\sqrt{\gamma}} \frac{\partial}{\partial \Psi}
[\alpha \sqrt{\gamma} (T^{\tilde{\Psi}}_{\tilde{\Psi}} + N^{\Psi} S_{\tilde{\Psi}})]
\nonumber \\
& \displaystyle  - \tilde{u} \frac{\partial \alpha}{\partial \Psi}
- \frac{\partial}{\partial \Psi} (\alpha N^j) S_{\tilde{j}}
+ \frac{1}{2}  \frac{\partial \gamma_{jk}}{\partial \Psi} T^{\tilde{j} \tilde{k}}.
\end{eqnarray}
The above equations cannot be used for 1D FFMD calculations.

\end{document}